\documentclass{IEEEtran}
\usepackage{cite}
\usepackage{amsmath,amssymb,amsfonts}
\usepackage{algorithmic}
\usepackage{graphicx}
\usepackage{textcomp}
\usepackage{color}
\usepackage{colortbl}

\usepackage{stfloats}
\usepackage{graphicx}
\usepackage{subcaption}
\begin{document}

\title{A Dual-Band 28/38-GHz Power Amplifier With Inter-Band Suppression in 22-nm FD-SOI CMOS for Multi-Standard mm-Wave 5G Communications}

\author{{Abbas Nasri, Alireza Yousefi, and Reza Nikandish$\rm ^*$}

\thanks{$\rm ^*$ Corresponding author}
\thanks{A. Nasri was with the School of Electrical and Electronic Engineering, University College Dublin, Dublin D04 V1W8, Ireland (e-mail: abbas.nasri@ieee.org).}
\thanks{A. Yousefi is with the Syntiant Corporation, Irvine, CA 92618 USA.}
\thanks{R. Nikandish is with the School of Electrical and Electronic Engineering, University College Dublin, Dublin D04 V1W8, Ireland (e-mail: reza.nikandish@ieee.org).}
}

\maketitle

\begin{abstract}
In this article, we present a dual-band 28/38-GHz power amplifier (PA) with inter-band suppression for millimeter-wave 5G communications. The dual-band operation is achieved using a center-tapped transformer network with an extra resonator which can provide optimum load impedance of the transistor in the two bands and synthesize a short-circuit between the two bands. This feature suppresses the PA signal emissions in the inter band, commonly allocated for other applications. A design procedure is developed for the proposed matching network including physical limits on the quality factor and the coupling coefficient of the transformer. The PA is designed using a 22-nm fully-depleted silicon-on-insulator (FD-SOI) CMOS process. The transistor stacking and a four-path transformer parallel-series power combining techniques are used to achieve high output power using the low-voltage process. The PA achieves simulated performance of 22.6/22.0\,dBm saturated output power, 19.8/20.0\,dBm output power at 1-dB gain compression, and 33/32\,\% maximum power-added efficiency (PAE) at 28/38\,GHz. The inter-band suppression is 6\,dB at 33\,GHz.
\end{abstract}

\begin{IEEEkeywords}
CMOS, dual-band, fifth generation (5G), fully-depleted silicon-on-insulator (FD-SOI), millimeter-wave, multi-band, power amplifier (PA), transformer.
\end{IEEEkeywords}

\section{Introduction}
 
\IEEEPARstart{M}{illimeter-Wave} (mm-wave) frequency bands are of paramount importance in the fifth generation (5G) and the future sixth generation (6G) networks where the broad spectrum availability can open up opportunities for new high-capacity communication applications \cite{rappaport-access2013, rappaport-access2019}. The development of innovative system and circuit architectures is essential to leverage exciting potentials of the mm-wave spectrum. A number of mm-wave bands are allocated for mobile communications, e.g., 28\,GHz, 38\,GHz, 60\,GHz, 140\,GHz, and several radio transceivers \cite{pang-jssc2019, lokhandwala2020high, tang2021design, wang-tmtt2022, Karakuzulu2023} and circuit components \cite{camarchia2020review, daneshgar2020high, sarkar201728, mortazavi2016integrated, dasgupta201926, Pashaeifar2022, wang2021broadband, callender2019band} operating in these bands are presented. There is an increasing quest for the concurrent multi-band circuits to develop universal and multi-standard products in smaller chip area and with lower cost \cite{lee2021millimeter, hu201928, ding202228, xu202128, del2019multi, hashemi-tmtt2002, nikandish2014, wang-tmtt2022, nikandish2016, celik-jssc2021, vigilante-jssc2018, chappidi2017frequency, mayeda2020highly}. 

There are three main approaches to realize the multi-band mm-wave circuits. In the most straightforward design method, multiple independent circuits each operating in one of the frequency band are used to realize the multi-band circuit. The operating band is selected by switches integrated in the circuits. This approach offers higher reliability and robustness and, as a result, is commonly used in commercial products. However, this leads to larger chip area, higher fabrication cost, and lower performance due to high insertion loss and low isolation of the switches in mm-wave bands. 

Another approach to multi-band operation is through broadband circuits covering multiple bands \cite{celik-jssc2021, vigilante-jssc2018, chappidi2017frequency, mayeda2020highly}. Examples include a 29--57-GHz class-AB power amplifier (PA) using a fourth-order matching network in 28-nm CMOS \cite{vigilante-jssc2018}, a 0.4--31.6\,GHz distributed PA in 22-nm fully-depleted silicon-on-insulator (FD-SOI) CMOS \cite{celik-jssc2021}, and a frequency reconfigurable dual-path 40--65-GHz PA in 130-nm SiGe BiCMOS \cite{chappidi2017frequency}. This approach offers a lower sensitivity to modeling inaccuracies and process variations, but suffers from two major issues. First, a broadband circuit usually requires complicated impedance matching networks which their high insertion loss due to the low quality factor of on-chip passive components can degrade the circuit performance. Second, the broadband PAs can transmit spurious signals in undesired communication bands, i.e., out-of-band emissions, while the broadband low-noise amplifiers (LNAs) can receive blocker signals and noise present in unintended bands. It is, therefore, essential to use additional filtering with such broadband circuits which leads to extra loss and higher system cost. 

In the third approach, the multi-band circuits are realized using dedicated multi-resonance circuit architectures \cite{hashemi-tmtt2002, nikandish2016, hu201928, ding202228, xu202128, lee2021millimeter, nikandish2014}. The design of these multi-band circuits requires special attention to the circuit functionality to develop impedance matching networks which can provide the required conditions at multiple frequencies. This approach has received increased interests recently, where a number of developments include a dual-band 28/38-GHz PA in 250-nm SiGe BiCMOS \cite{ding202228}, a tri-band 28/37/39-GHz Doherty PA with reconfigurable matching networks in 130-nm SiGe BiCMOS \cite{hu201928}, a dual-band 28/38-GHz PA in 22-nm FD-SOI CMOS \cite{xu202128}, and a dual-band 27/33-GHz amplifier with transformer-feedback neutralization of the gate-drain capacitance in 100-nm GaAs pHEMT \cite{nikandish2016}. The main challenge of this approach is the development of multi-band circuits which can provide a good performance in the presence of practical limitations of the process, e.g., low quality factor of passive elements, parasitic capacitances, modeling inaccuracies, and process variations.

In this article, we present a dual-band 28/38-GHz PA using a transformer-based impedance matching network. The transformer network, which provides the impedance matching and differential-to-single ended transformation, is \textit{center-tapped with an extra resonator}. This resonator enables the transformer network to operate at two frequency bands and provide the \textit{inter-band suppression}. This attenuates the PA out-of-band emissions in the frequency bands allocated for other communication applications. 

The article is organized as follows. In Section II, we present the dual-band matching network architecture, principles of operation, and a detailed analysis of the impacts of circuit imperfections on its performance. In Section III, we discuss the circuit design of a dual-band PA using the developed transformer network realized in 22-nm FD-SOI CMOS process. The post-layout simulation results of the PA are presented in Section IV and the conclusions are discussed in Section V. 

\section{Dual-Band Transformer Network}\label{MN_DB}

\subsection{Proposed Network}
\begin{figure*}[!t]
  \begin{center}
  \includegraphics[width=1.9\columnwidth]{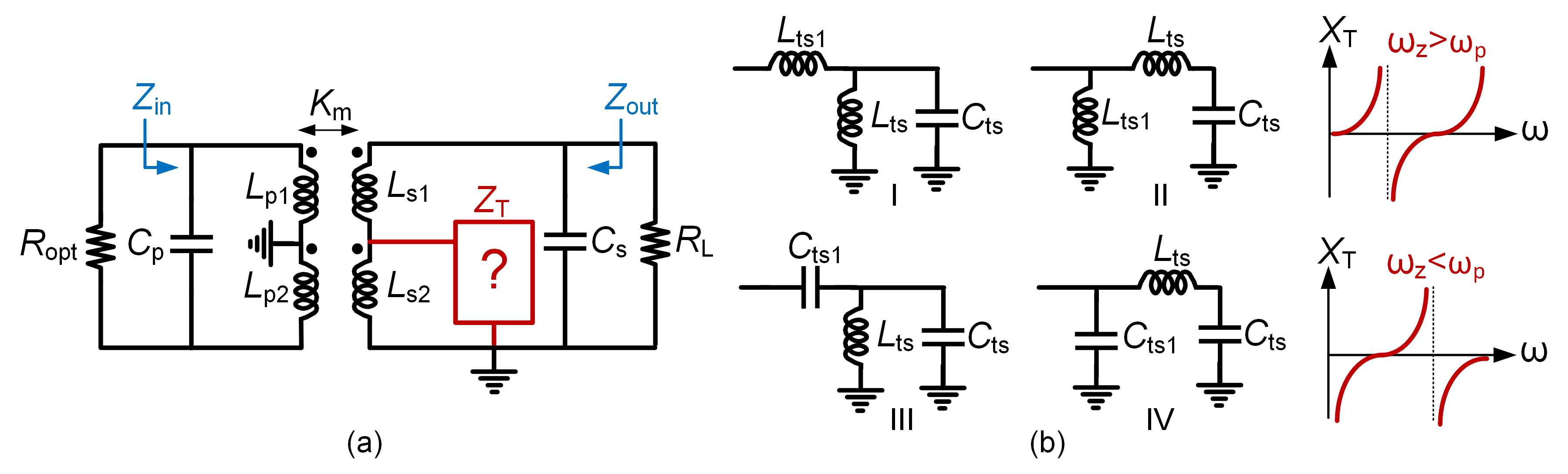}
  \caption{(a) Dual-band transformer network with center-tap resonator. (b) Four possible three-element circuits as the resonator.}\label{MN_Proposed}
  \end{center}
\end{figure*}

The proposed transformer network for dual-band impedance matching and inter-band suppression is shown in Fig. \ref{MN_Proposed}(a). The secondary winding of the transformer is center-tapped with a reactive resonator network. We will discuss the selection criteria for this resonator circuit later. 
The source impedance of the transformer network is modeled by the optimum load resistance of the transistor $R_{opt}$ in parallel with a capacitance $C_p$ which should be absorbed into the transformer network. The capacitance $C_p$ comprises the output-referred parasitic capacitance of the transistor and the primary parasitic capacitance of the transformer. 
 The load impedance of the network includes the load resistance $R_L$ in parallel with a capacitance $C_s$, which comprises secondary parasitic capacitance of the transformer and parasitic capacitance of the output signal pad. 
 
 The transformer turn ratio is selected such that it transforms the load impedance $R_L$ to the optimum resistance $R_{opt}$, while the transformer inductances absorb the parasitic capacitances $C_p$ and $C_s$. These conditions should be concurrently satisfied at the lower and upper frequencies, $\omega_L$ and $\omega_H$, to realize a dual-band impedance matching network

\begin{equation}
    \label{Zin}
    Z_{in}(j \omega_L) = Z_{in}(j \omega_H) = R_{opt}.
\end{equation}
It should be noted that the resistive part of the input impedance can be set to the optimum resistance by the proper selection of the transformer turns ratio as 
\begin{equation}
    \label{turn_ratio}
    n \approx \sqrt{\frac{R_{opt}}{R_L}}.
\end{equation}
However, it is challenging to satisfy the reactive part conditions. \textit{A conventional transformer cannot absorb the parasitic capacitances in two frequency bands}. A double-tuned transformer network which can provide two peaks in its frequency response is useful as interstage network of broadband amplifiers and resonator of oscillators \cite{vigilante-jssc2018, mazzanti2018second, bevilacqua2020doubly}. This network has disadvantages of loss, imbalanced magnitudes in the two peaks, and tuning complexity as the ratio of the two peak frequencies is mainly controlled by the transformer coupling coefficient. 
Therefore, we propose the transformer network of Fig. \ref{MN_Proposed}(a) with the center-tap impedance of $Z_T(j\omega)$ which should provide three conditions: 
\begin{enumerate}
    \item At the lower frequency $\omega_L$, the resonator should operate as an open-circuit. The transformer's primary and secondary inductances are $L_p = L_{p1} + L_{p2}$ and $L_s = L_{s1} + L_{s2}$, and resonate with the parasitic capacitances.
    \item At the higher frequency $\omega_H$, the resonator should be equivalent with a short-circuit to effectively decrease the inductances of the transformer to $L_{p1}$ and $L_{s1}$. This enables the transformer to absorb the parasitic capacitances at the higher frequency. 
    \item At the inter-band frequency $\omega_{SC}$, it should provide a proper reactance which, along with the transformer, creates a short-circuit at \textit{output port of the transformer}. This realizes the expected inter-band suppression. 
\end{enumerate}

The discussed conditions can be summarized as follows.
\begin{equation}\label{Z_T}
 \begin{cases}
 Z_{T}(j\omega_{L})=\infty \ \\ Z_{T}(j\omega_{H})=0 \ \\ Z_{out}(j\omega_{SC})=0 
 \end{cases}
\end{equation}
The first two conditions are only dependent on the resonator network's elements, while the third condition is also related to the transformer's elements. We develop a design approach for the dual-band matching network. 


In the dual-band transformer circuit of Fig. \ref{MN_Proposed}(a), several circuit structures can be envisioned as the resonator network. We consider the four resonator circuits shown in Fig. \ref{MN_Proposed}(b) as the possible realizable circuits using three reactive elements. These circuits have a zero and a pole in their impedance frequency response. The requirements of the resonator network given in (\ref{Z_T}) indicate that the circuit should ideally have a pole at $\omega_L$ and a zero at $\omega_H$. This condition suggests to use the network I or II with $\omega_z > \omega_p$. However, these networks include two inductors which can lead to higher loss and larger chip area compared to the networks III and IV with only one inductor. We can use the network III or IV which their pole is set at $\omega_L$ and the impedance at $\omega_H$ is practically small enough such that it can be approximated by zero. This condition can be considered as $|Z_T(j\omega_H)| \ll R_L$ or, for the sake of future derivations, as follows
\begin{equation}
    \label{ZT_wh}
    |Z_T(j\omega_H)| = {\delta} R_L,
\end{equation}
where $\delta$ is a constant which can be set in the typical range of 0.01--0.1 depending on the required accuracy and circuit element values. For the resonator network, we select the network III and derive the design criteria.

\subsection{Design Approach}
The impedance of the resonator network III in Fig. \ref{MN_Proposed}(b) can be derived as
\begin{equation}\label{Zins}
 Z_{T}(j\omega)= \frac{1-\omega^2 L_{ts}(C_{ts}+C_{t1s})}{1-\omega^2 L_{ts}C_{ts}} \frac{1}{j \omega C_{ts1}}.
\end{equation}
The resonator network should have a pole at $\omega_L$, leading to
\begin{equation}
    \label{wL_condition}
    \omega_L^2 L_{ts} C_{ts} = 1.
\end{equation}
The resonator network should meet the condition (\ref{ZT_wh}) at $\omega_H$ which using (\ref{Zins}) and (\ref{wL_condition}) can be derived as
\begin{equation}
    \label{wH_condition}
    \frac{\left( 1 + \frac{C_{ts1}}{C_{ts}} \right) \omega_H^2 - \omega_L^2}{\omega_H^2 - \omega_L^2}  \frac{1}{\omega_H C_{ts1} R_L} = \delta.
\end{equation}
This can be solved for $C_{ts1}$ as follows
\begin{equation}
    \label{whH_condition2}
    C_{ts1} = \frac{C_{ts}}{{\delta} \omega_HC_{ts}R_L - \alpha},
\end{equation}
where $\alpha = \omega_H^2/(\omega_H^2 - \omega_L^2)$. This indicates that the parameter $\delta$ should satisfy the condition $\delta > \alpha/(\omega_H C_{ts} R_L)$ and cannot be selected arbitrarily small. Therefore, the minimum theoretical impedance of the resonator in (\ref{ZT_wh}) is derived as 
\begin{equation}
    \label{ZT_wh_min}
    |Z_T(j\omega_H)|_{\rm min} = \frac{\omega_H^2}{\omega_H^2 - \omega_L^2} \frac{1}{\omega_H C_{ts}}.
\end{equation}




The short-circuit frequency of the transformer network should meet the condition $\omega_L < \omega_{SC} < \omega_H$. We can consider it as the mid logarithmic distance between the lower and higher frequencies, leading to
\begin{equation}
    \label{SC_freq}
    \frac{\omega_{SC}}{\omega_L} = \frac{\omega_H}{\omega_{SC}} \Rightarrow  \omega_{SC}= \sqrt{\omega_{L} \omega_{H}}. 
\end{equation}
The impedance of the resonator network at the short-circuit frequency can be derived using (\ref{Zins}), (\ref{wL_condition}), and (\ref{SC_freq}) as
\begin{equation}
    \label{ZT_sc}
    Z_{T}(j\omega_{SC}) = \frac{\left( 1 + \frac{C_{ts1}}{C_{ts}} \right) \omega_H - \omega_L}{\omega_H - \omega_L}  \frac{1}{j\omega_{SC} C_{ts1} },
\end{equation}
which is equivalent with a capacitance given by
\begin{equation}
    \label{Csc}
    C_{sc} = \frac{(\omega_H - \omega_L)C_{ts1}} {\left( 1 + \frac{C_{ts1}}{C_{ts}} \right) \omega_H - \omega_L}.
\end{equation}
This capacitance can generate a resonance with the transformer to produce a short-circuit under certain conditions. The output impedance of the transformer network should be derived to apply the inter-band suppression condition, $Z_{out}(j\omega_{SC}) = 0$. Using the circuit of Fig. \ref{MN_Proposed} with the resonator network III, assuming $L_{p1} = L_{p2} = \frac{1}{2} L_p$ and $L_{s1} = L_{s2} = \frac{1}{2} L_s$, and setting $Z_{out}(j\omega_{SC}) = 0$, it can be shown that the following condition should be satisfied
\begin{equation}
    \label{ZSC_condition}
    Z_{T}(j\omega_{SC}) + \frac{1}{4} j\omega_{SC} L_s = 0. 
\end{equation}
Using (\ref{ZT_sc}) and (\ref{Csc}), this condition can be derived as
\begin{equation}
    \label{wsc_resonance}
    \frac{1}{4} \omega_{SC}^2 L_{s} C_{sc} = 1.
\end{equation}

The resonator network can be designed using the conditions given by (\ref{wL_condition}), (\ref{ZT_wh_min}), and (\ref{wsc_resonance}). There are extra degrees of freedom in this system of equations which provide more design flexibility. The transformer network can be designed using the approximate input-referred equivalent circuit of the transformer (assuming $K_m \approx 1$) comprising a parallel RLC circuit with the elements
\begin{equation}
    \label{Lin}
    L_{in} \approx L_p 
\end{equation}
\begin{equation}
    \label{Cin}
    C_{in} \approx C_p + \frac{1}{n^2}C_s
\end{equation}
\begin{equation}
    \label{Rin}
    R_{in} \approx n^2 R_L.
\end{equation}
The network is designed such that $ L_{in}$ and $C_{in}$ resonate at the lower frequency $\omega_L$, while $R_{in}=R_{opt}$. The resonator network effectively reduces the inductances to $L_{p1}$ and $L_{s1}$ at the higher frequency $\omega_H$, enabling the network to resonate at both $\omega_L$ and $\omega_H$. An important advantage of this network is that the transformer coupling coefficient is not a design parameter and the transformer layout can be optimized for the maximum coupling. This is in contrary with the conventional double-tuned transform network which usually should be designed with a low coupling coefficient to achieve broadband response at the cost of a lower gain \cite{vigilante-jssc2018, mazzanti2018second, bevilacqua2020doubly}.

We use the transducer power gain as the performance metric for the transformer network such that the impact of losses can also be evaluated \cite{nikandish2021broadband}. This is defined as the ratio of the output power delivered to the load to the available source power and can be derived as
\begin{equation}
    \label{GT}
    G_T(\omega) = \frac{4 R_{opt} R_L |Z_{21}|^2}{|(Z_{11} + R_{opt})(Z_{22} + R_L) - Z_{12}Z_{21}|^2},
\end{equation}
where $Z_{ij}$ denotes the impedance parameter of the transformer two-port network. 

\begin{figure}[!t]
  \begin{center}
  \includegraphics[width=0.9\columnwidth]{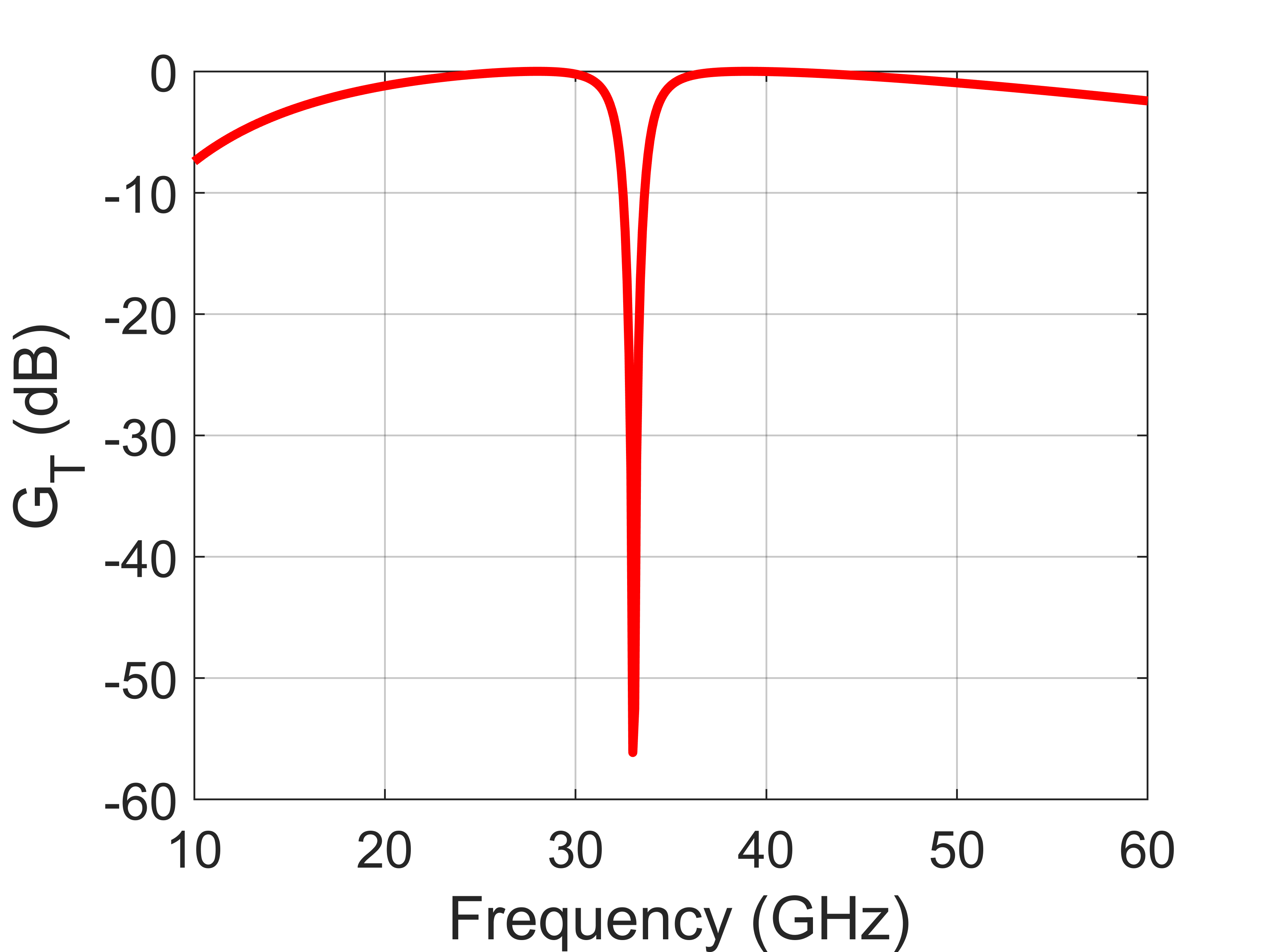}
  \caption{Transfer function of the dual-band 28/38-GHz transformer network realized using ideal elements.}\label{GT_IdealQ}
  \end{center}
\end{figure}

In Fig. \ref{GT_IdealQ}, $G_{T}(\omega)$ of the dual-band transformer network is shown. The network is realized using ideal lossless elements and perfect transformer coupling ($K_m = 1$). The network is designed for the lower frequency of 28\,GHz, the upper frequency of 38\,GHz, and the short-circuit frequency of $f_{SC} = \sqrt{f_Lf_H} \approx$\,33\,GHz.

\begin{figure}[!t]
  \begin{center}
  \includegraphics[width=0.9\columnwidth]{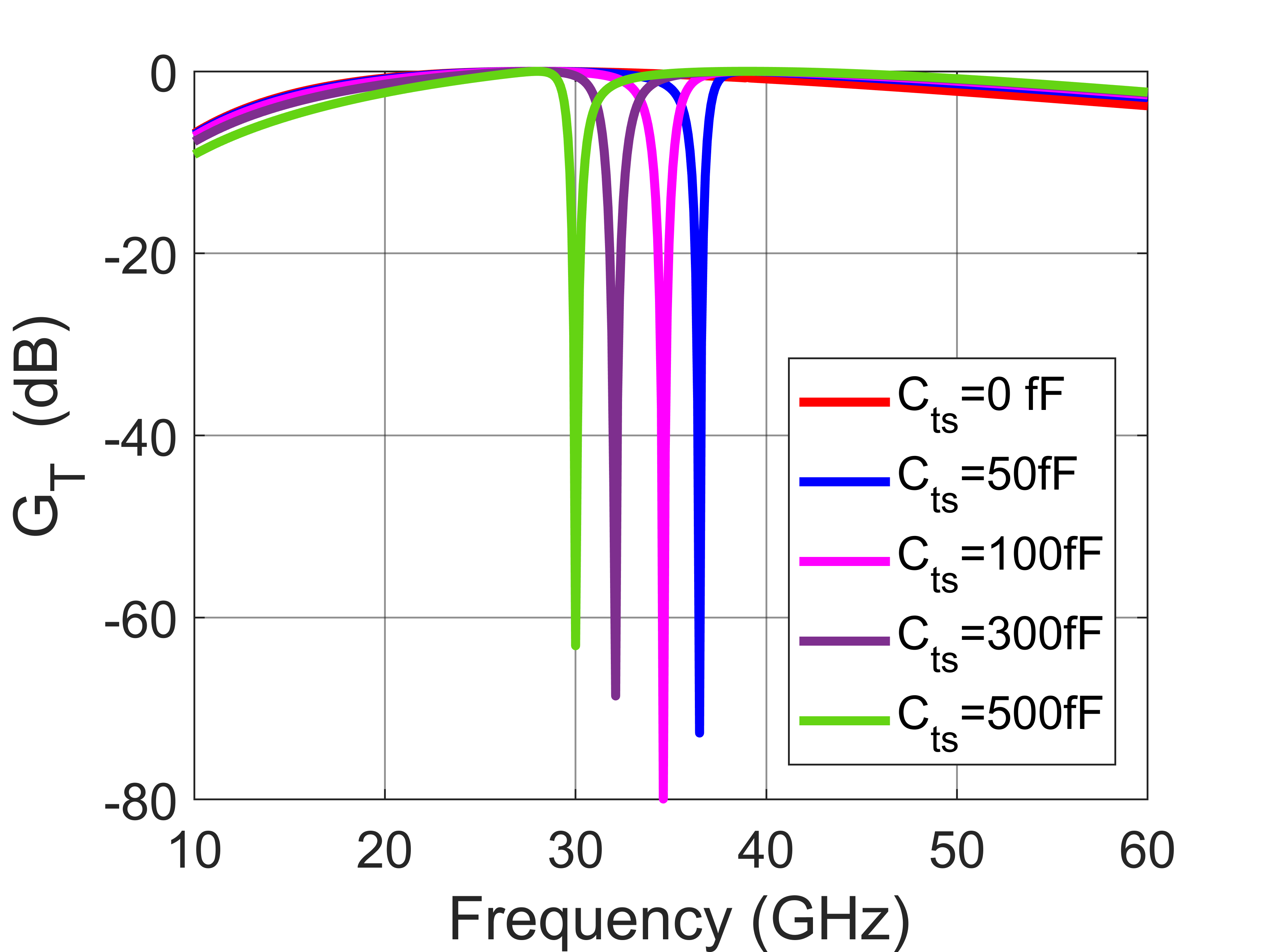}
  \caption{ Transfer function of the dual-band transformer with swept ${C_{ts}}$. The notch frequency can be controlled by ${C_{ts}}$.}\label{GT_IdealQ_Cts}
  \end{center}
\end{figure}

In Fig. \ref{GT_IdealQ_Cts}, the impact of changing the capacitance $C_{ts}$ on the frequency response is shown. This indicates that the short-circuit frequency can be controlled by using a digitally controlled or switched $C_{ts}$. The network has no notch for $C_{ts}=0$, while the notch is shifted toward the lower frequencies by increasing $C_{ts}$. 


\subsection{Impact of Transformer Loss}
The transformer loss arises from the limited quality factor of inductors, due to metal and substrate losses, and the imperfect transformer coupling. It is assumed that the primary and secondary inductors of the transformer have the quality factor of $Q_{\rm XFMR}$. This can be modeled by resistances $r_p = \omega L_p/Q_{\rm XFMR}$ and $r_s = \omega L_s/Q_{\rm XFMR}$ in series with $L_p$ and $L_s$, respectively. For the inductor of the resonator network with a quality factor of $Q_{T}$, the loss can be similarly modeled as a resistance $r_{ts} = \omega L_{ts}/Q_{ T}$ in series with the inductor $L_{ts}$. 

\begin{figure}[!t]
  \begin{center}
  \includegraphics[width=0.9\columnwidth]{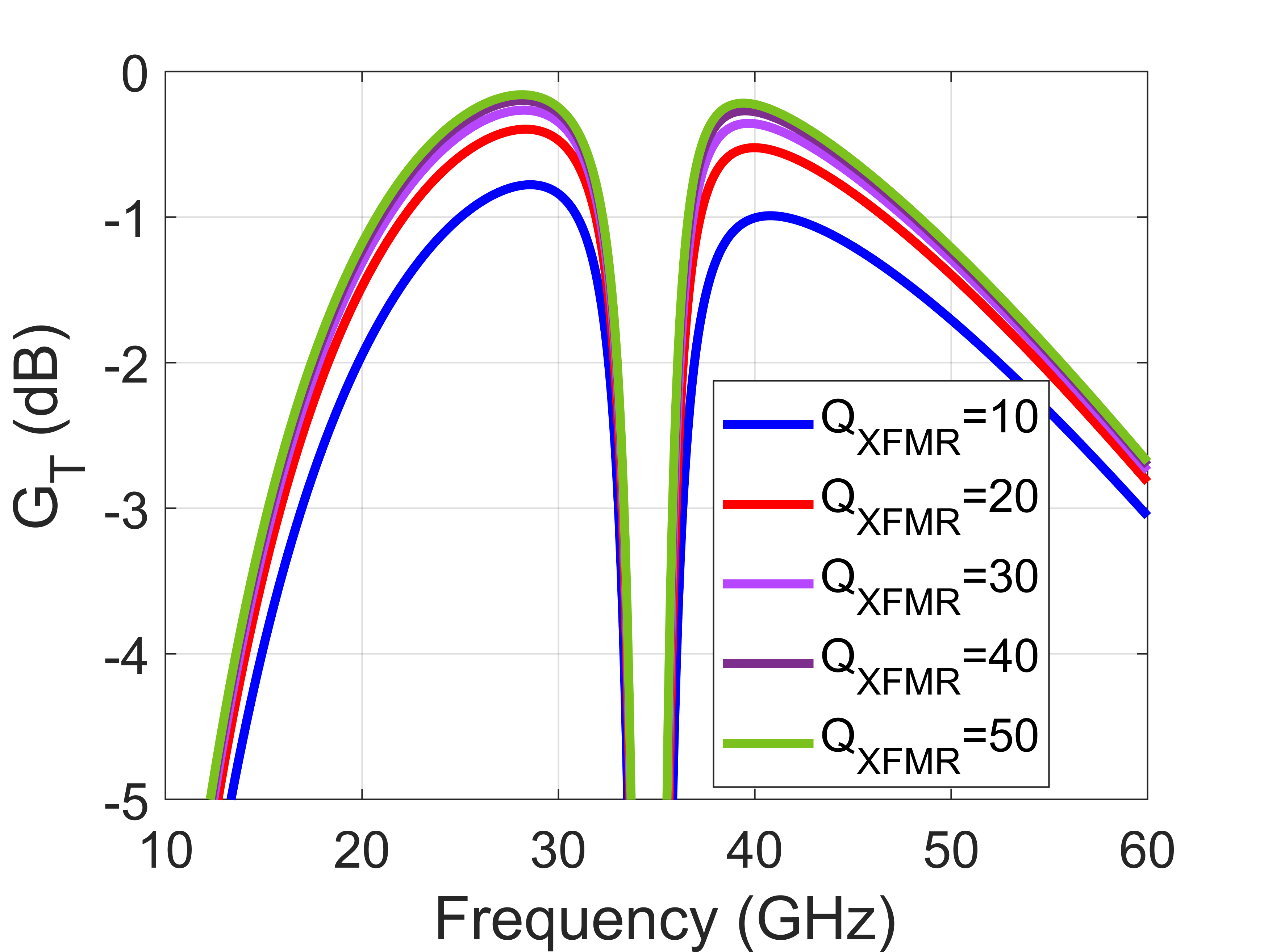}
  \caption{Transfer function of the dual-band transformer for different values of the transformer quality factor (assuming $K_m=1$ and $Q_T=30$).}\label{GT_realQ_Trans}
  \end{center}
\end{figure}

In Fig. \ref{GT_realQ_Trans}, the impact of transformer quality factor on the transfer function is illustrated. The insertion loss in the two pass bands increases by lowering $Q_{\rm XFMR}$. For high $Q_{\rm XFMR}$, the insertion loss is limited by the resonator network quality factor $Q_T$ which is assumed to be 30 in this simulation. A transformer coupling coefficient of 1 is used to focus on the effects of the transformer quality factor. As will be discussed in Section III, the transformer physical structure is an hexagonal spiral to comply with design rules of the process. This limits its quality factor to about 20--25. 

\begin{figure}[!t]
  \begin{center}
  \includegraphics[width=0.9\columnwidth]{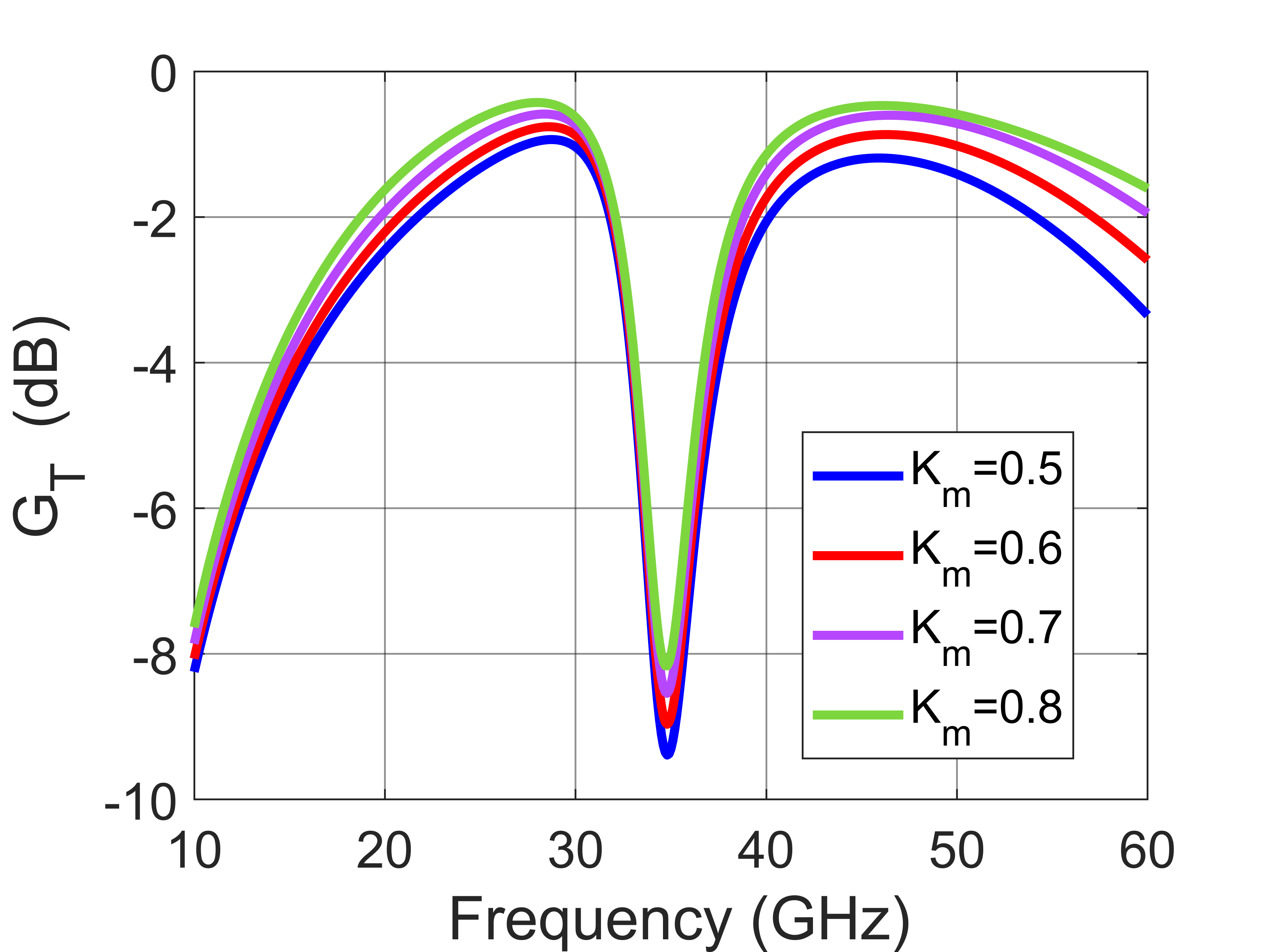}
  \caption{Transfer function of the dual-band transformer with swept coupling coefficient $K_{m}$ (assuming $Q_T=30$). }\label{GT_RealQ_Tuning_Km}
  \end{center}
\end{figure}

In Fig. \ref{GT_RealQ_Tuning_Km}, the transfer function versus frequency is shown for different transformer coupling coefficients $K_{m}$. The insertion loss increase when $K_{m}$ is reduced from 0.8 to 0.5. The impact is more significant in the upper band. 

\subsection{Impact of Resonator Loss}

\begin{figure}[!t]
  \begin{center}
  \includegraphics[width=0.9\columnwidth]{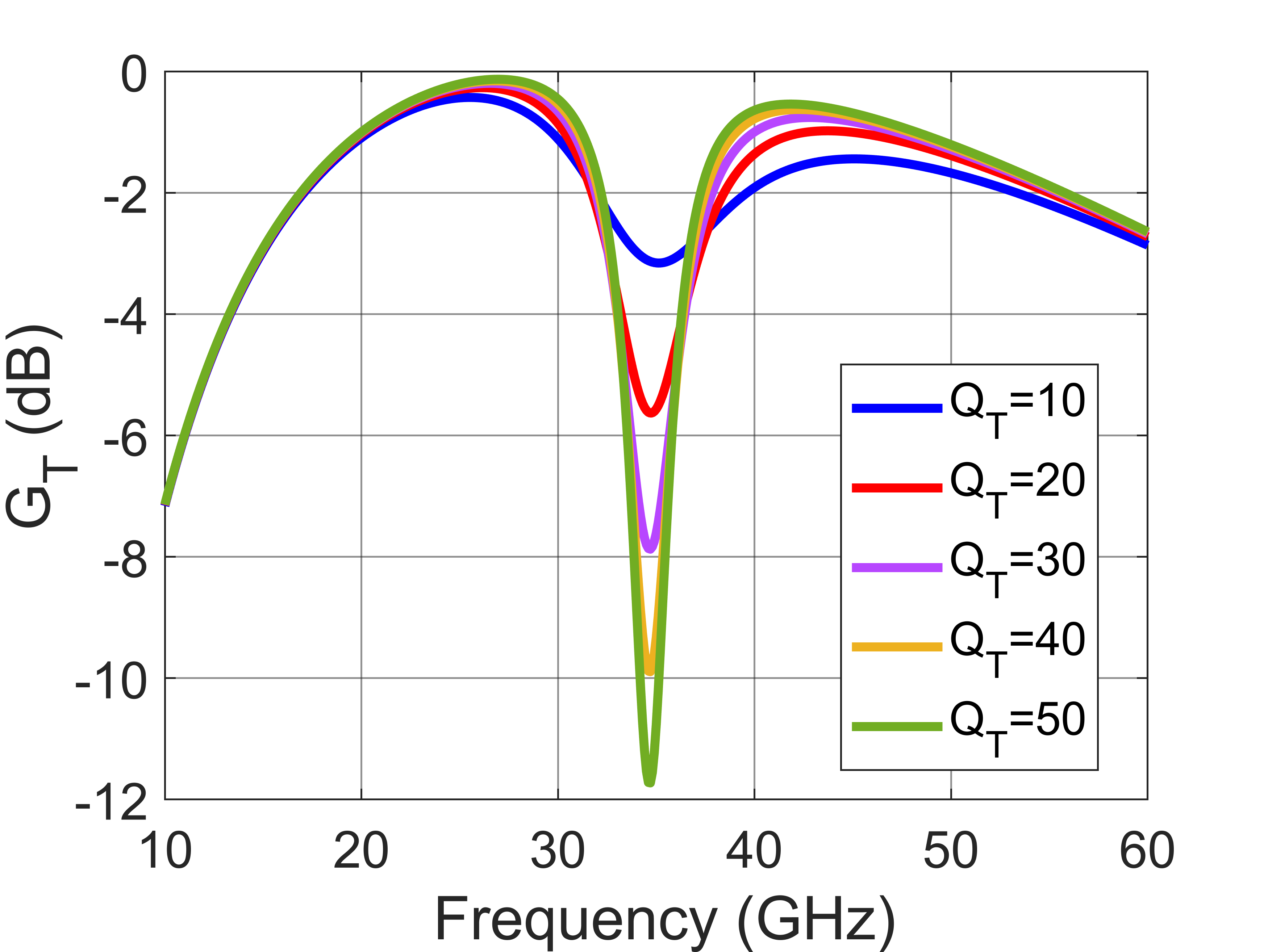}
  \caption{ Transfer function of the dual-band transformer network for different values of the inductor quality factor $Q_T$ in the resonator network.}\label{GT_RealQ_Tuning}
  \end{center}
\end{figure}

The resonator loss is mainly caused by the limited quality factor of the inductor $L_{ts}$. This is modeled by a resistance $r_{ts} = \omega L_{ts}/Q_{ T}$ in series with the inductor $L_{ts}$ in the network III of Fig. \ref{MN_Proposed}(b). The effects of $Q_T$ on the transfer function of the transformer network are evaluated in Fig. \ref{GT_RealQ_Tuning}. The resonator quality factor has significant impact on the suppression at the notch frequency. For example, to achieve 10\,dB suppression, the quality factor of the resonator inductor should be at least 40 which is difficult to reach in this low-resistivity substrate process. Fortunately, the resonator inductor can be implemented as a straight transmission line, unlike the hexagonal spiral shape transformer, which features a higher quality factor compared to the transformer.

\section{Power Amplifier Circuit Design}
\begin{figure*}[!t]
  \begin{center}
  \includegraphics[width=2.0\columnwidth]{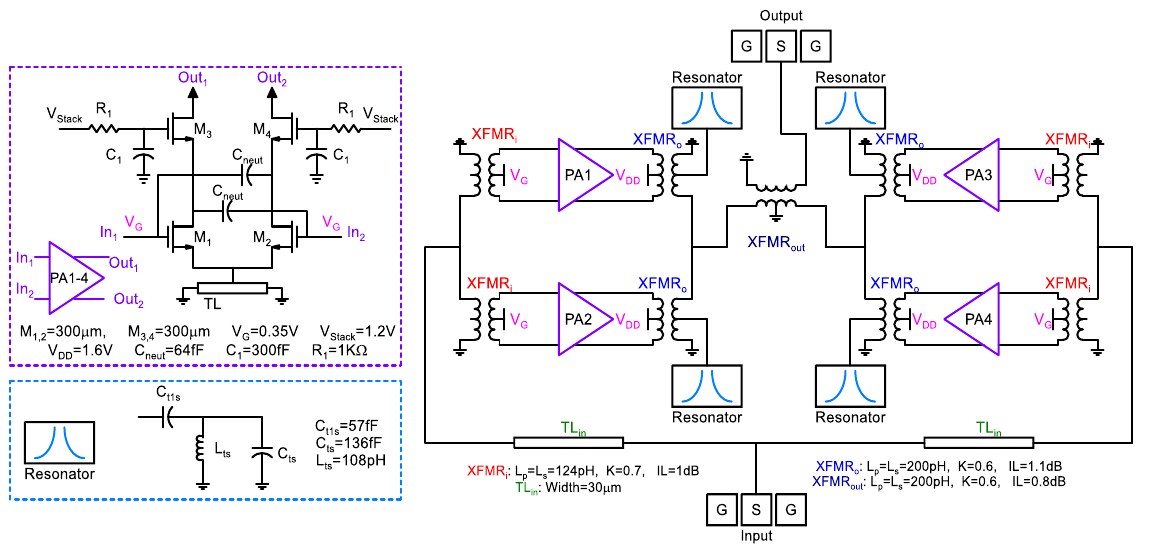}
  \caption{Schematic of the dual-band power amplifier.}\label{PA_Proposed}
  \end{center}
\end{figure*}

The dual-band PA architecture and circuit details are shown in Fig. \ref{PA_Proposed}. The PA comprises four power cells (PA1--4) combined in the parallel-series configuration. The power cells are matched to 50\,$\Omega$ using the dual-band transformer network center-tapped with the resonator ($\rm XFMR_o$). The parallel combining of output signals from two transformers $\rm XFMR_o$ also transforms the impedance level to 25\,$\Omega$ which is then returned to 50\,$\Omega$ through the series combining by the transformer $\rm XFMR_{out}$. The input power is divided between the power cells using the transformers $\rm XFMR_i$ and the transmission lines $\rm TL_{in}$. The power cells are realized using differential double-stacked structure shown in Fig. \ref{PA_Proposed}.  We discuss details of the PA circuit design. 


\subsection{FD-SOI CMOS Process}
\begin{figure}[!t]
  \begin{center}
  \includegraphics[width=0.9\columnwidth]{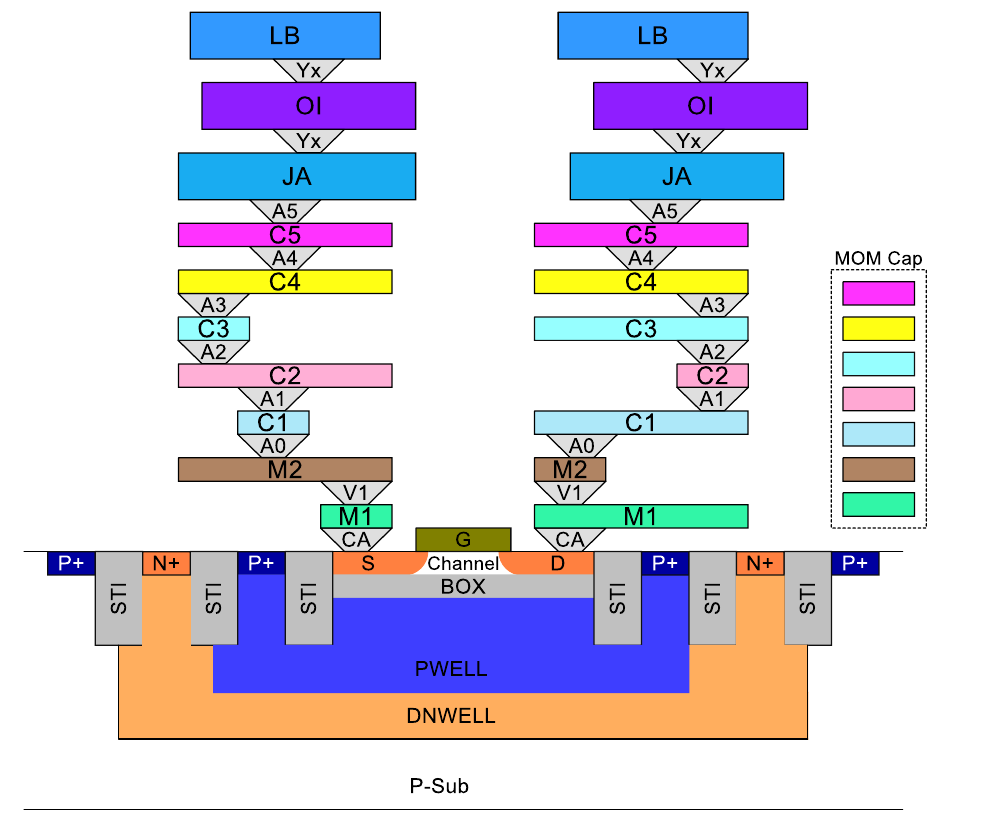}
  \caption{The 22-nm FD-SOI CMOS process structure.}\label{Technology}
  \end{center}
\end{figure}

The PA is implemented in the Global Foundries 22-nm fully-depleted silicon-on-insulator (FD-SOI) CMOS process (22FDX). The process structure is shown in Fig. \ref{Technology} which includes one Aluminum top thick metal layer, 2 thick and 7 thin metal Copper layers. The substrate has a low resistivity (7\,$\rm \Omega.cm$) unlike the conventional SOI processes. The transformers in the PA circuit are implemented on the two top copper layers (IA and OI) which have a lower resistivity compared to the top Aluminum layer (LB). The circuit capacitors are realized as the metal-oxide-metal (MOM) capacitors using stacked thin metal layers (M1-2 and C1-5). 

The process offers multiple types of transistors with different threshold and breakdown voltages. Super-low threshold voltage (SLVT) transistors are used in this design to benefit from their high transconductance $g_m$, unity current gain frequency $f_T$, and unity unilateral power gain frequency $f_{\rm max}$. The SLVT NMOS transistor features 0.25\,V threshold voltage, 0.8\,V nominal supply voltage, $f_T$ of 350\,GHz, and $f_{max}$ of 370 GHz. The process also provides body bias possibility to adjust threshold voltage of transistors which has not been used in this design.



\subsection{Stacked Power Cell}
The circuit schematic of the power cells is shown in Fig. \ref{PA_Proposed}. The low breakdown voltage of the thin-oxide transistors limits their supply voltage and output power capability. Therefore, two transistors are stacked to increase the maximum supply voltage from nominal 0.8\,V to 1.6\,V. In the double-stacked amplifier, the gate node of the top transistors $M_{3,4}$ is biased through a large resistor $R_1$ operating roughly as open-circuit for the RF signal, unlike the Cascode amplifier in which the gate of top transistors is RF grounded. An accurately designed capacitor $C_1$ is included in the gate of top transistor to control the gate-source voltage swing of this device. This results in higher output power and efficiency in the stacked amplifier. The input transistors $M_{1,2}$ are biased at the class-AB to improve their gain and power-added efficiency (PAE). 

\begin{figure}[!t]
  \begin{center}
  \includegraphics[width=\columnwidth]{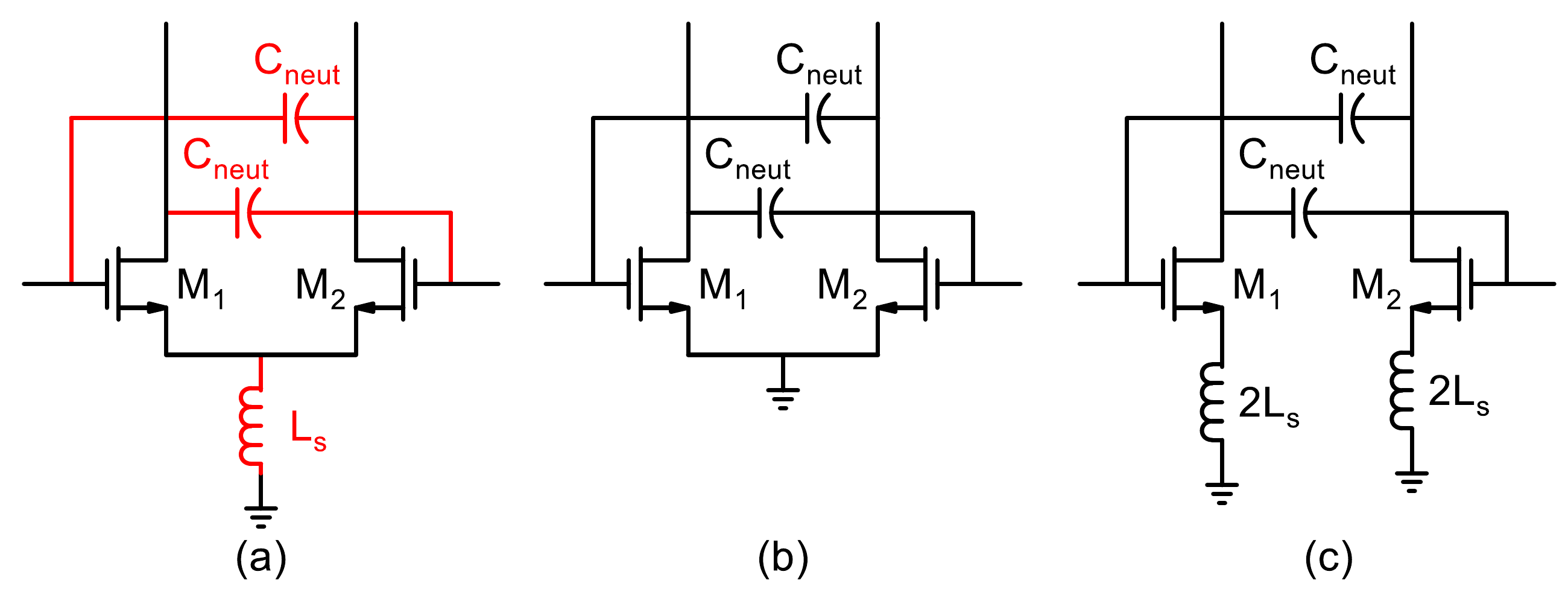}
  \caption{(a) Power cell differential amplifier with stability network, (b) differential-mode circuit, (c) common-mode circuit.}\label{Fig_Stablity_Sch}
  \end{center}
\end{figure}

\begin{figure}[!t]
  \begin{center}
  \includegraphics[width=\columnwidth]{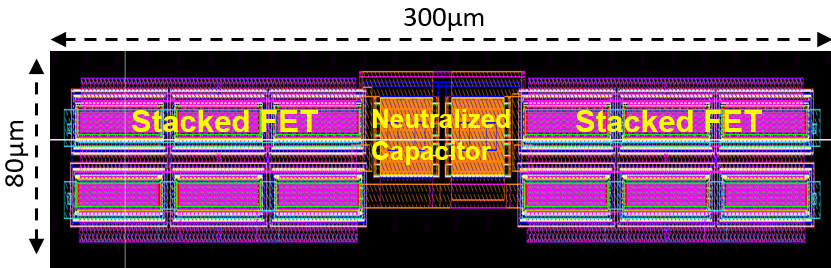}
  \caption{Layout of the power cell.}\label{PA_Layout}
  \end{center}
\end{figure}

The power cell is realized as a differential amplifier, shown in Fig. \ref{Fig_Stablity_Sch}(a), where the cross-connected neutralization capacitors $C_{neut}$ are used to cancel the gate-drain capacitors and achieve unconditional stability. Furthermore, an inductor $L_{s}$ is placed in the common node of the input transistors pair to improve the common-mode stability. This inductor has no effect in the differential mode, as shown in Fig. \ref{Fig_Stablity_Sch}(b), while it reduces gain of the transistors in the common mode, as can be inferred from Fig. \ref{Fig_Stablity_Sch}(c). Layout of the power cell is shown in Fig. \ref{PA_Layout}.


\subsection{Size of Transistors}
\begin{figure}[!t]\centering
\begin{minipage}{0.24\textwidth}\centering
  \includegraphics[width=\textwidth]{ 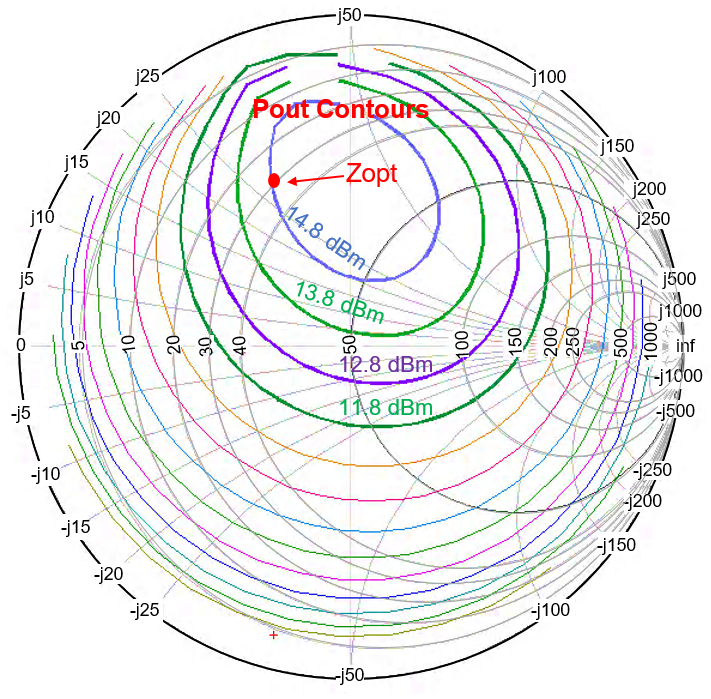}
    \subcaption{\label{Fig_Load_Pull_28}}
\end{minipage}
\begin{minipage}{0.24\textwidth}\centering
  \includegraphics[width=\textwidth]{ 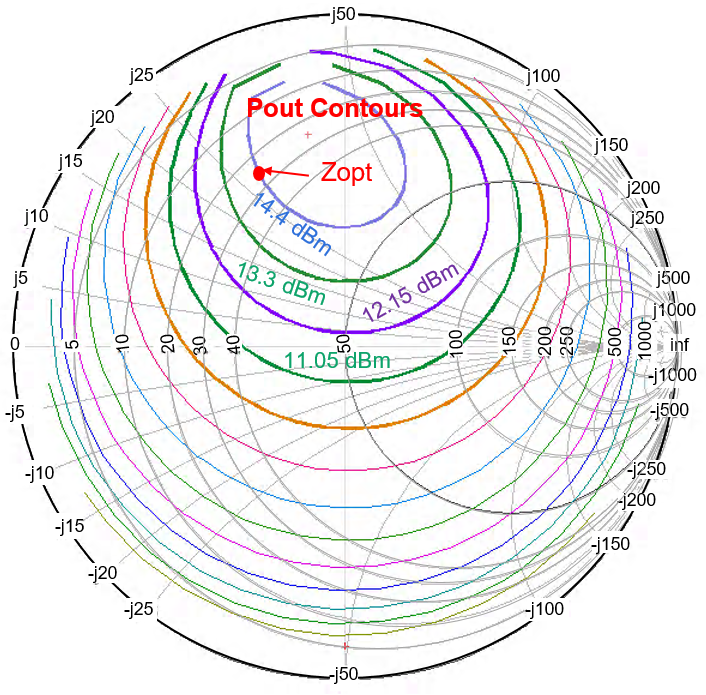}
    \subcaption{\label{Fig_Load_Pull_38}}
  \end{minipage}
  \caption{The load-pull simulation result for output power at 1-dB gain compression of the power cell: (a) 28\,GHz, (b) 38\,GHz. The optimum output power is 14.8/14.4\,dBm at 28/38\,GHz. 
  }\label{Fig_Load_Pull}
\end{figure}

The gate length of all transistors is set at the minimum length of the process, 20\,nm. to achieve the best RF performance. The width of transistors is selected based on the output power (14\,dBm at 1-dB gain compression) and optimum load resistance (close to 50\,$\Omega$) requirements for each power cell. The width is determined using the load-pull and source-pull simulations performed on the power cell. The load-pull simulation results at the two frequencies 28\,GHz and 38\,GHz are shown in Fig. \ref{Fig_Load_Pull}. The optimum output power is 14.8/14.4\,dBm at 28/38\,GHz.



\subsection{Output Power Combiner}

\begin{figure}[!t]
  \begin{center}
  \includegraphics[width=0.9\columnwidth]{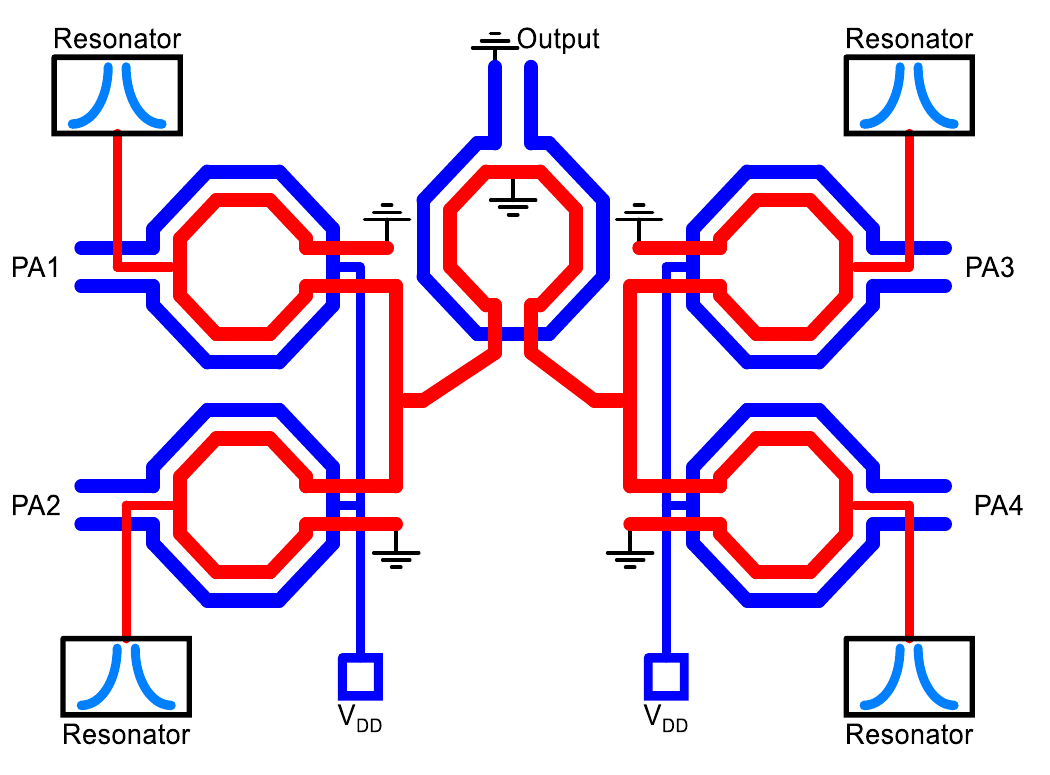}
  \caption{Layout of the output power combiner and impedance matching network.}\label{Output_Combiner}
  \end{center}
\end{figure}

The output power combiner which also transforms the load resistance to the optimum load resistance of transistors is realized using four dual-band transformer networks and a transformer as output power combiner. The real part of the optimum load impedance of the power cells is about 50\,$\Omega$. Thus, the dual-band transformer $\rm XFMR_o$ has a 1:1 turn ratio. The parallel combining of two power cells transforms the resistance level to 25\,$\Omega$ which is then converted to 50\,$\Omega$ using the series combining. This allows the output transformer $\rm XFMR_{out}$ also be realized as a 1:1 transformer. This realizes a parallel-series power combining which can ideally convert the 14\,dBm output power of the power cell to 20\,dBm output power. The output network layout is shown in Fig. \ref{Output_Combiner}.

\begin{figure}[!t]\centering
\begin{minipage}{0.24\textwidth}\centering
  \includegraphics[width=\textwidth]{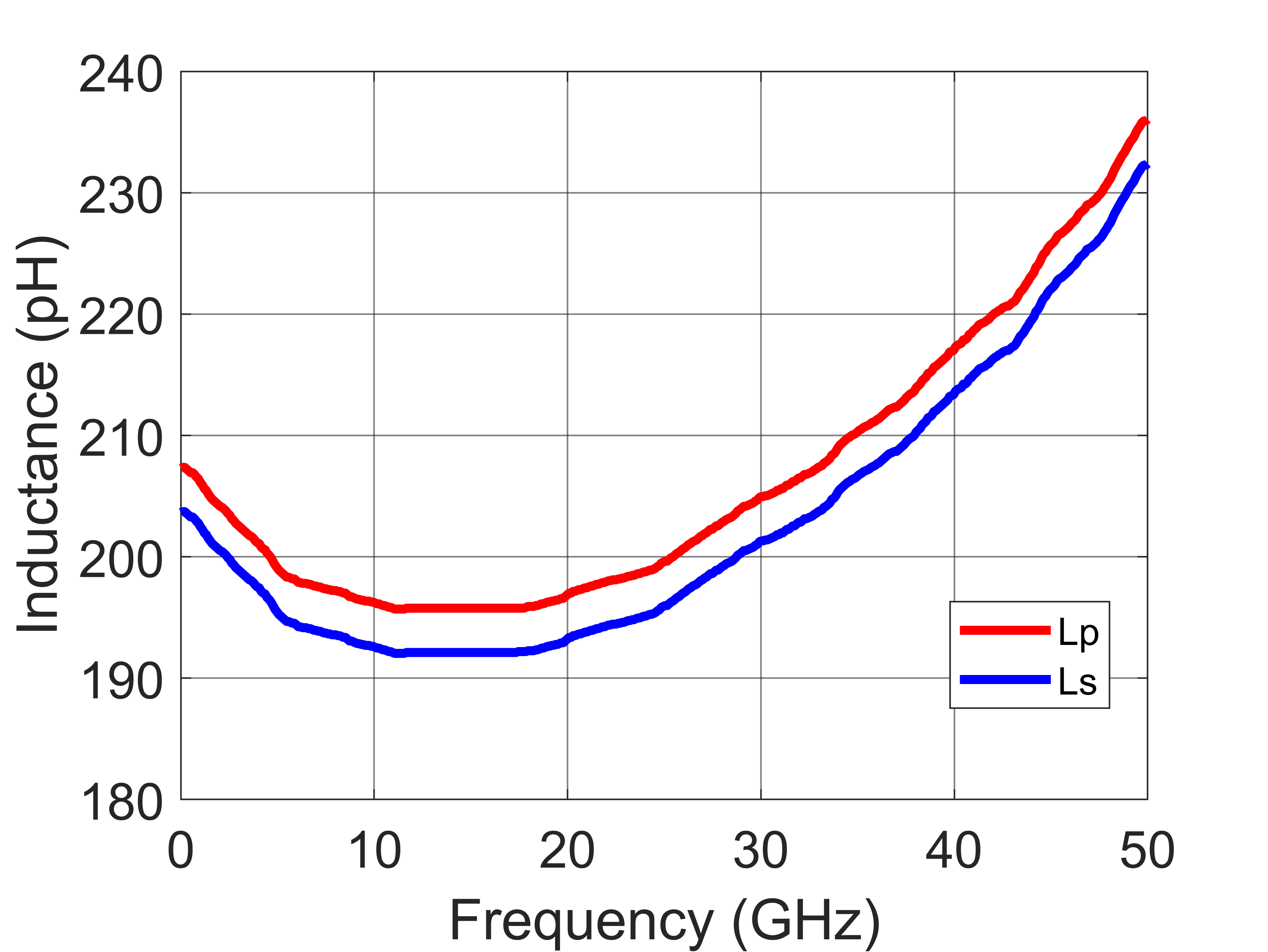}
    \subcaption{\label{L_trans}}
\end{minipage}
\begin{minipage}{0.24\textwidth}\centering
  \includegraphics[width=\textwidth]{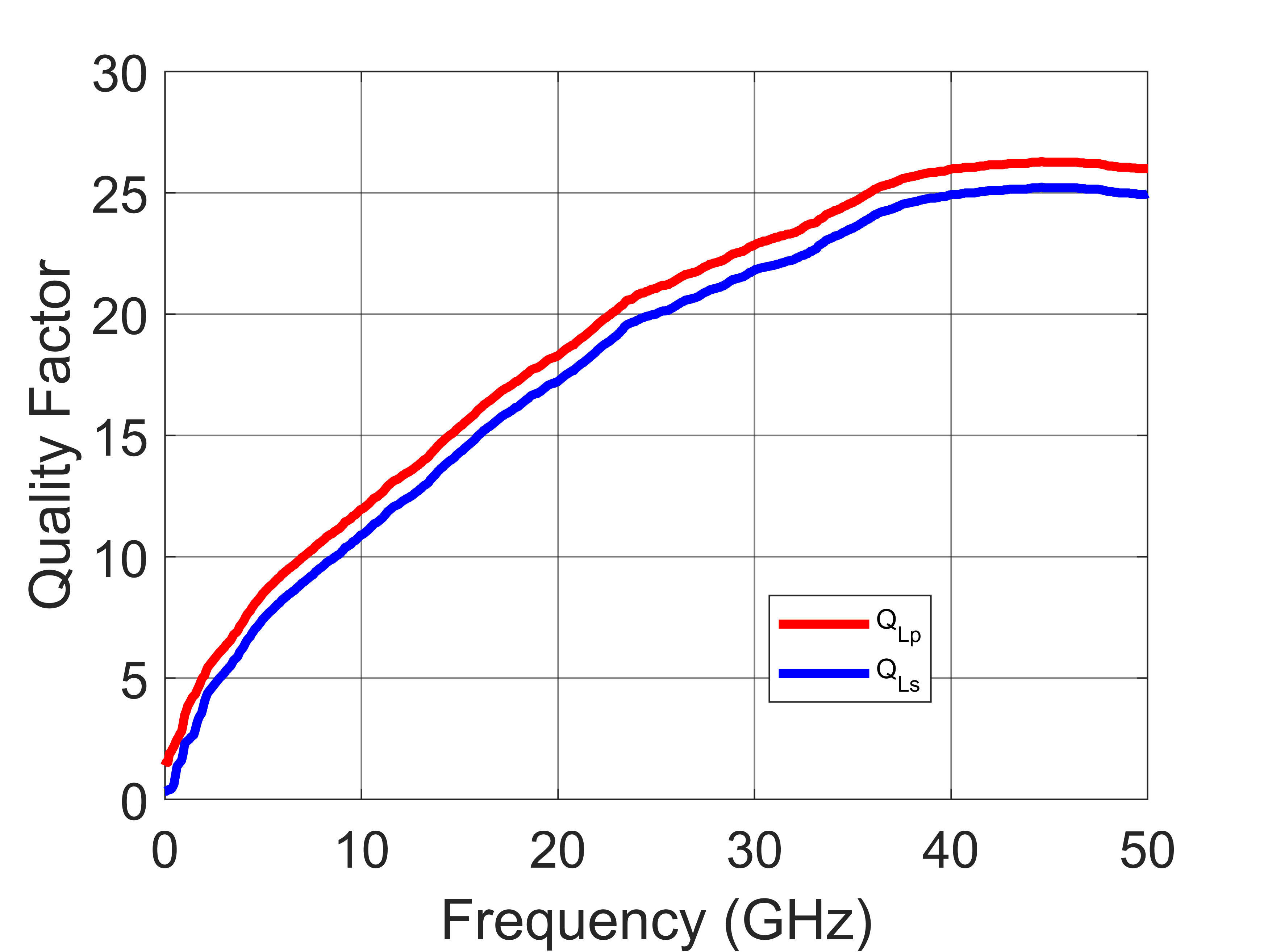}
    \subcaption{\label{Q_Trans}}
  \end{minipage}
  \caption{Extracted (a) inductance and (b) quality factor of the output transformer $\rm XFMR_o$.}\label{Trans_Result}
\end{figure}

\begin{figure}[!t]
  \centering
  \begin{subfigure}[b]{0.49\linewidth}
    \includegraphics[width=\linewidth]{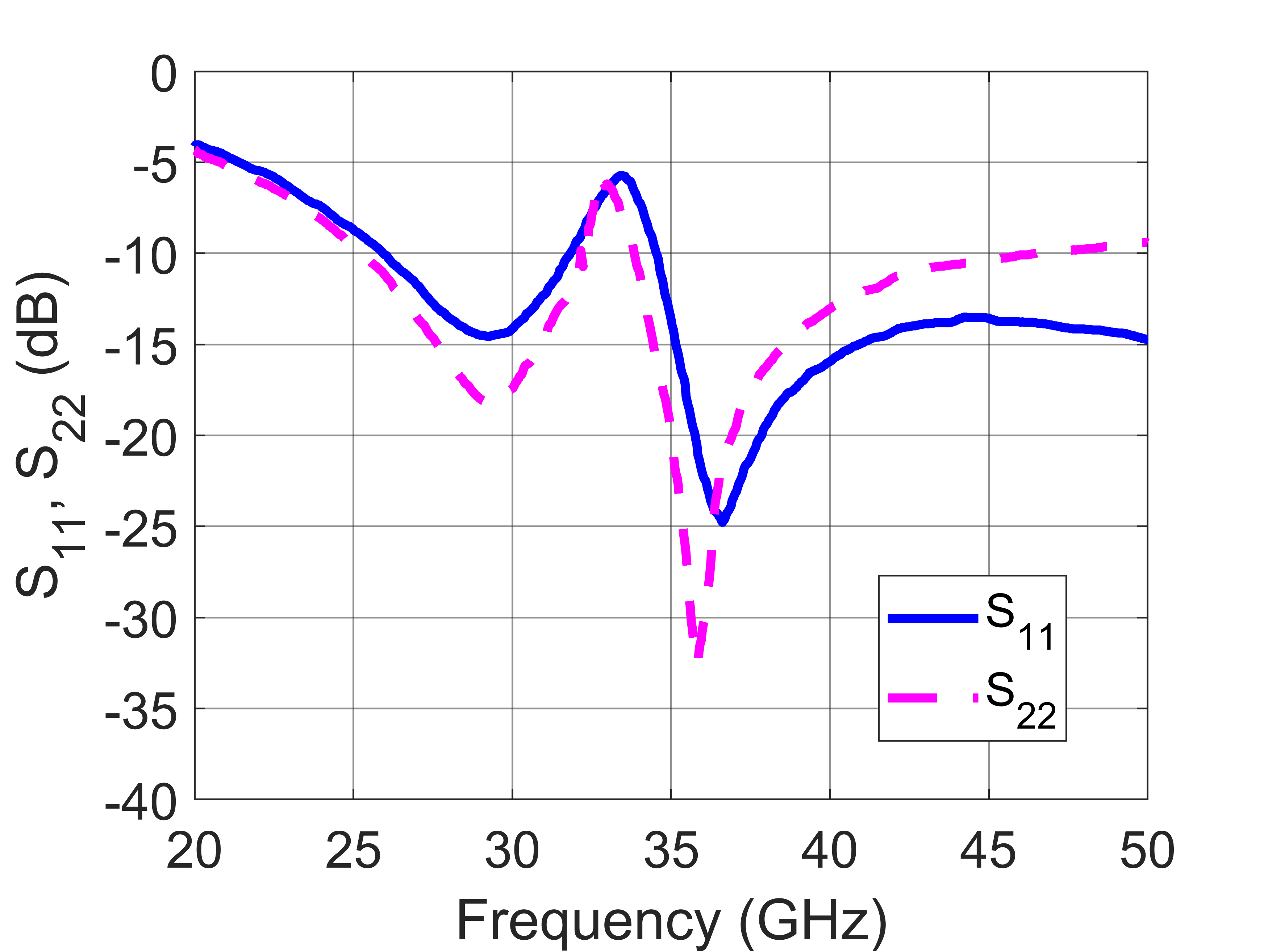}
    \caption{}
  \end{subfigure}
  \begin{subfigure}[b]{0.49\linewidth}
    \includegraphics[width=\linewidth]{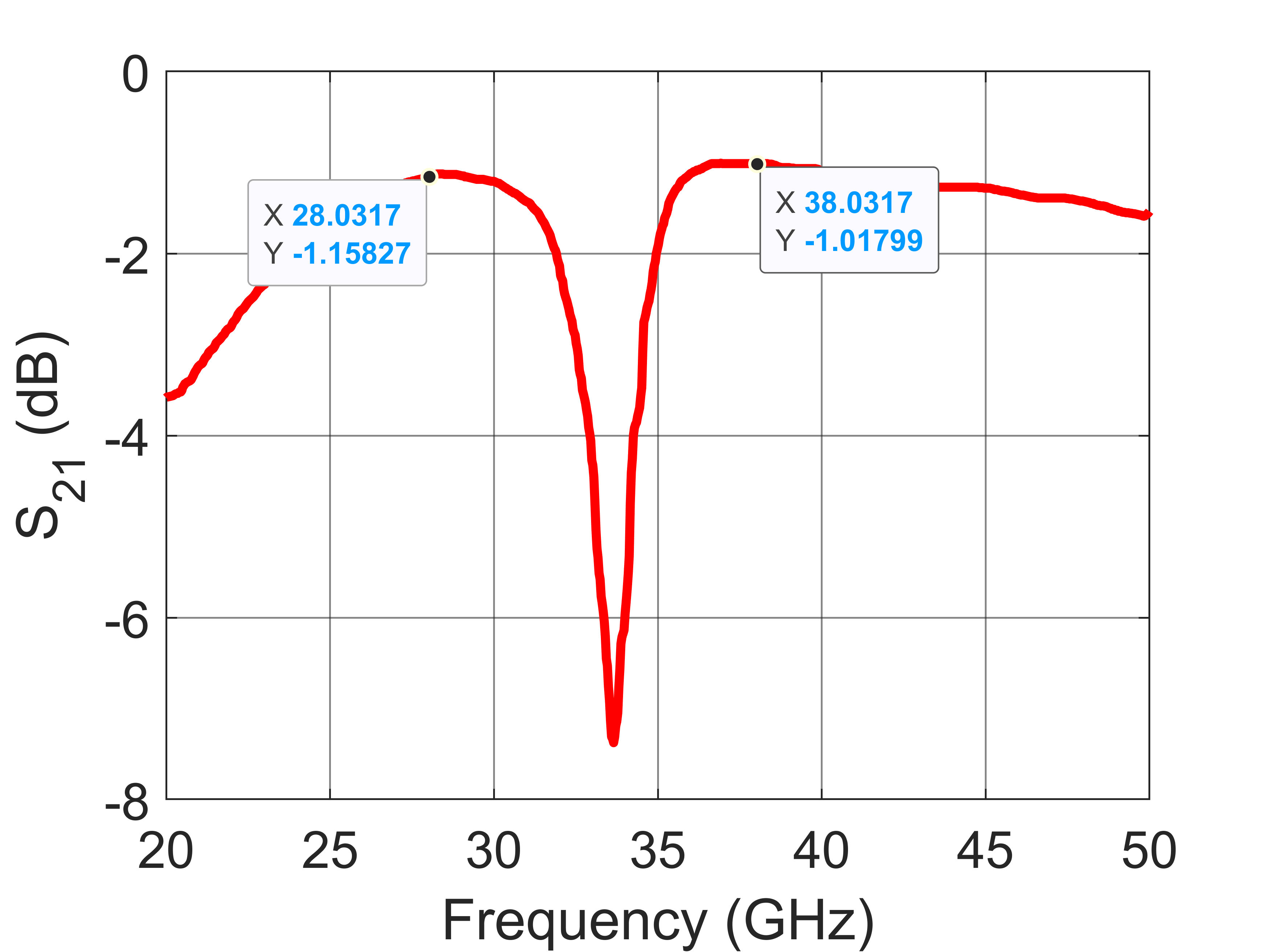}
    \caption{}
  \end{subfigure}
  \caption{ Simulated scattering parameters of the dual-band transformer with the center-tap resonator network: (a) $S_{11}$ and $S_{22}$, (b) $S_{21}$.}
  \label{S_Parm_DB_MN}
\end{figure}

The power combiner performance is simulated using the EMX planar 3D electromagnetic (EM) simulator. The extracted inductance and quality factor of the dual-band transformers are shown in Fig. \ref{Trans_Result}. The inductance is about 200\,pH and the quality factor is 22--25. Simulated scattering parameters of the dual-band transformer network are shown in Fig. \ref{S_Parm_DB_MN}. The insertion loss is about 1\,dB, the inter-band suppression is 6\,dB, and the input/output return losses are higher than 15\,dB.


\subsection{Input Power Splitter}

\begin{figure}[!t]
  \begin{center}
  \includegraphics[width=0.6\columnwidth]{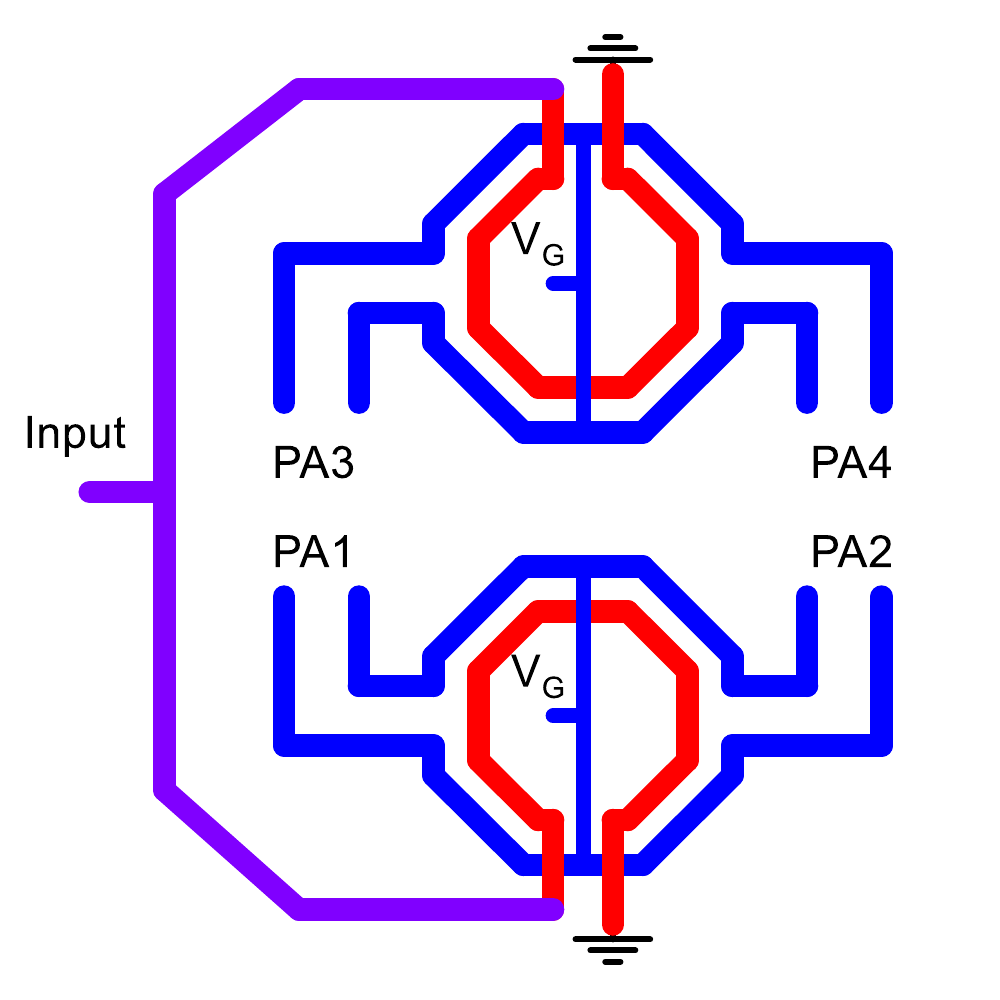}
  \caption{Layout of the input power splitter and impedance matching network.}\label{Input_Trans}
  \end{center}
\end{figure}

The input power splitter also serving as the input impedance matching network is shown in Fig. \ref{Input_Trans}. This network comprises two single-to-differential transformers and two transmission lines. The input impedance of the transformers should be 100\,$\Omega$ to provide an input impedance 50\,$\Omega$ with the two-way power splitter. The input transformers match the 100\,$\Omega$ resistance to the optimum source impedance of the transistors. 

\begin{figure}[!t]
  \begin{center}
  \includegraphics[width=0.8\columnwidth]{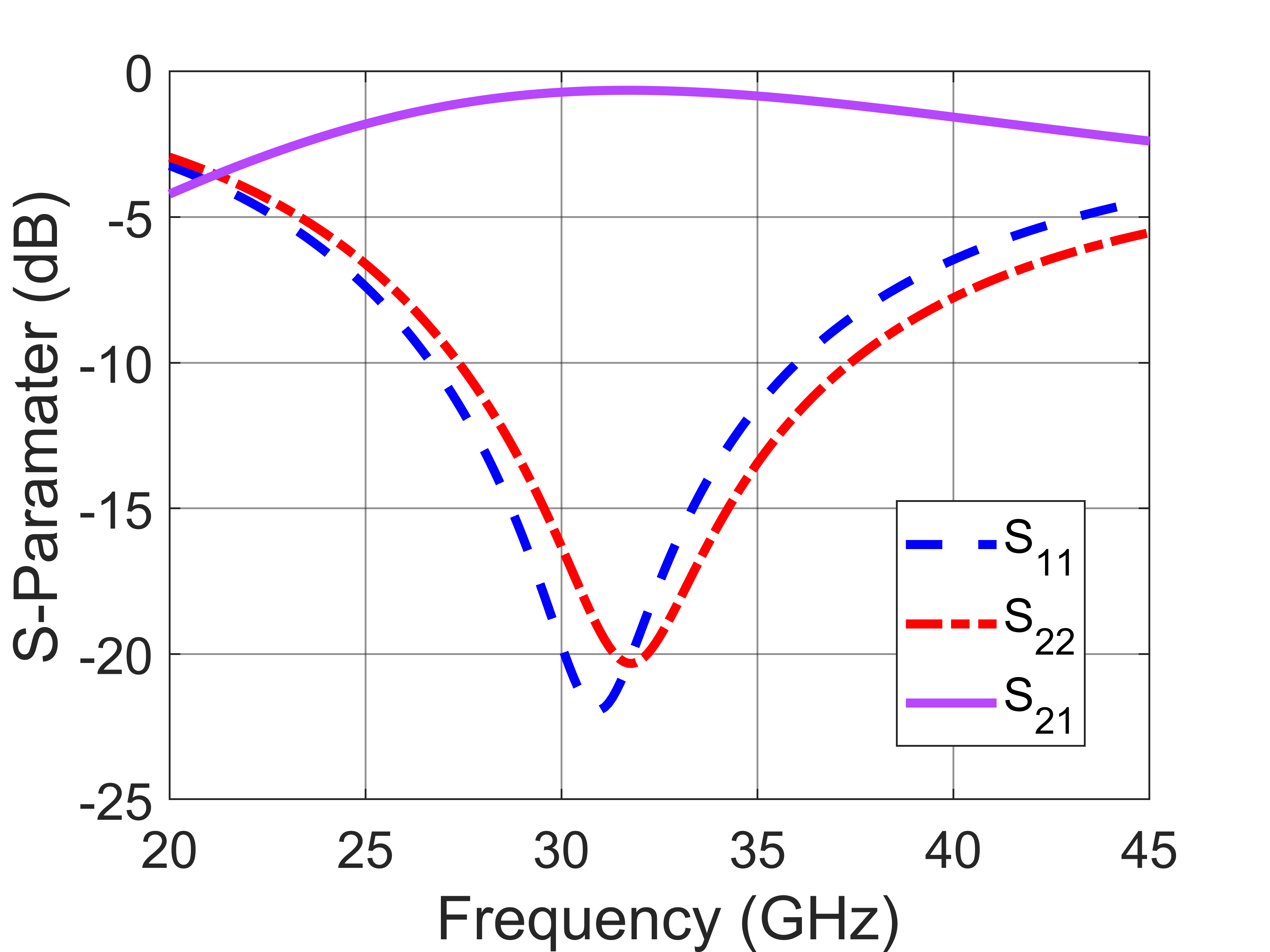}
  \caption{Simulated scattering parameters of the input network. Insertion loss is lower than 1.2\,dB in the target band.}\label{SP_TRin}
  \end{center}
\end{figure}

The input network is designed with a broad bandwidth that covers the lower and upper bands. Scattering parameters of the input network are shown in Fig. \ref{SP_TRin}, which indicates an insertion loss lower than 1.2\,dB in the target bands. 


\section{Power Amplifier Simulation Results}
\begin{figure}[!t]
  \begin{center}
  \includegraphics[width=2.5in]{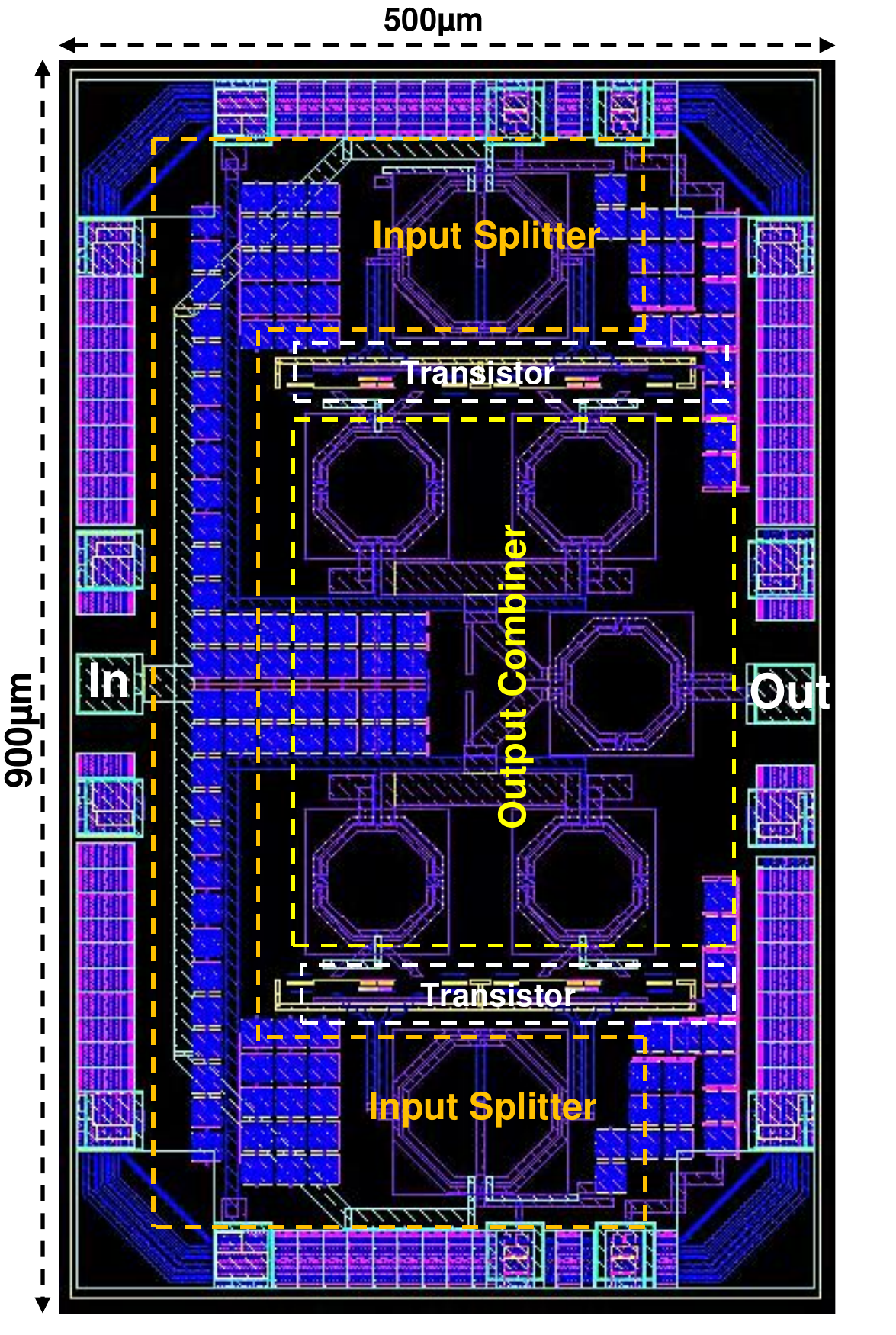}
  \caption{Layout of the PA in 22-nm FD-SOI process.}\label{Layout_PA}
  \end{center}
\end{figure}

The layout of the PA is shown in Fig. \ref{Layout_PA}, where the chip measures 0.5\,mm $\times$ 0.9\,mm. The PA is biased at the supply voltage of $\rm V_{DD}=1.6\,V$, the gate bias voltage of $\rm V_{G}=0.35\,V$ for the input transistors, and $\rm V_{stack}=1.2\,V$ for the stack devices. The PA consumes $\rm 80\,mA$ drain current in the quiescent condition. The PA small-signal and large-signal simulations are performed using the 22-nm FD-SOI process design kit (PDK) for active devices and full electromagnetic simulations of the passive devices. 

\subsection{Small-Signal Simulations}
\begin{figure}[t!]
  \centering
  \begin{subfigure}[b]{0.9\linewidth}
    \includegraphics[width=\linewidth]{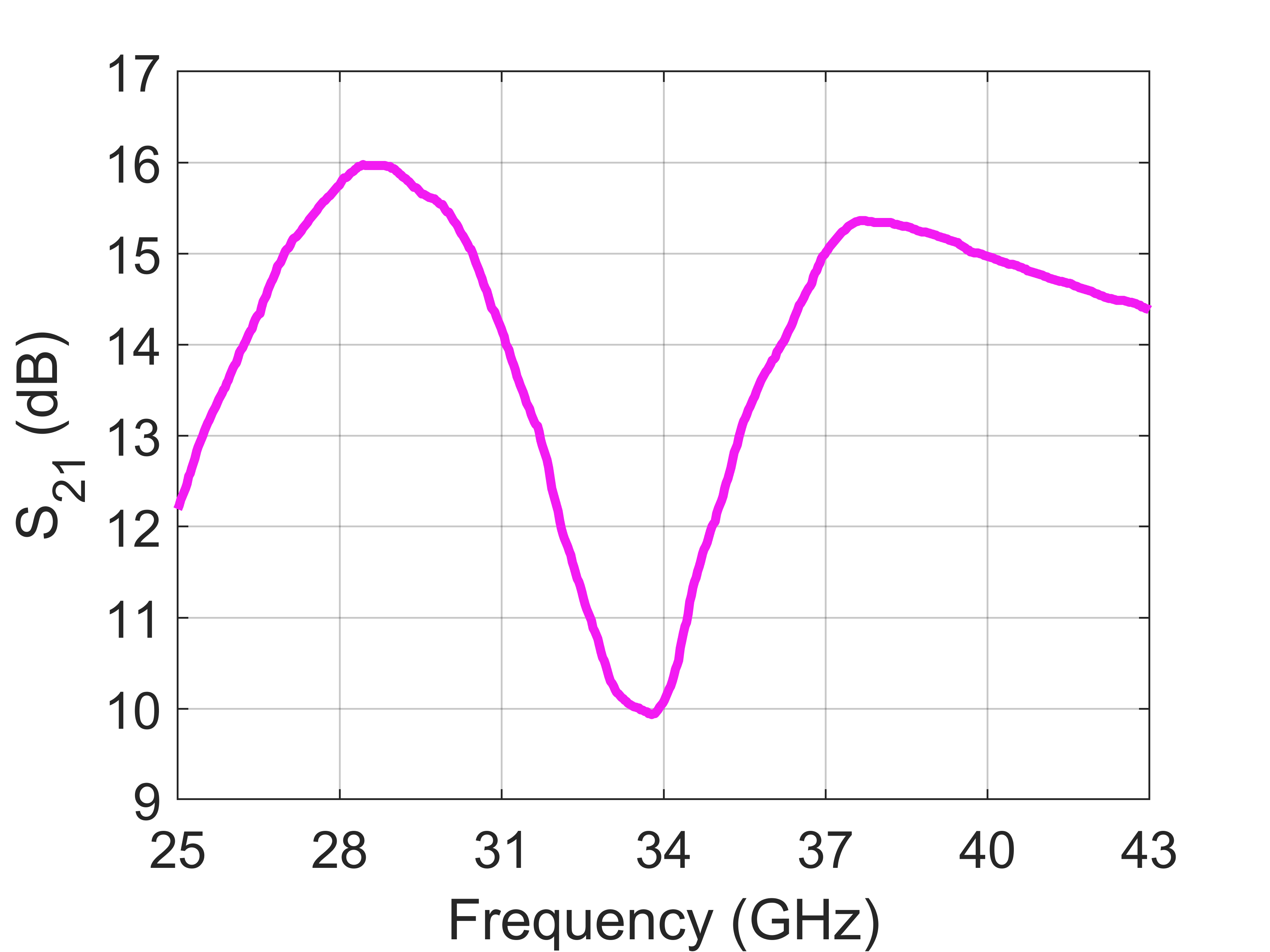}
    \caption{}
  \end{subfigure}
  \begin{subfigure}[b]{0.9\linewidth}
    \includegraphics[width=\linewidth]{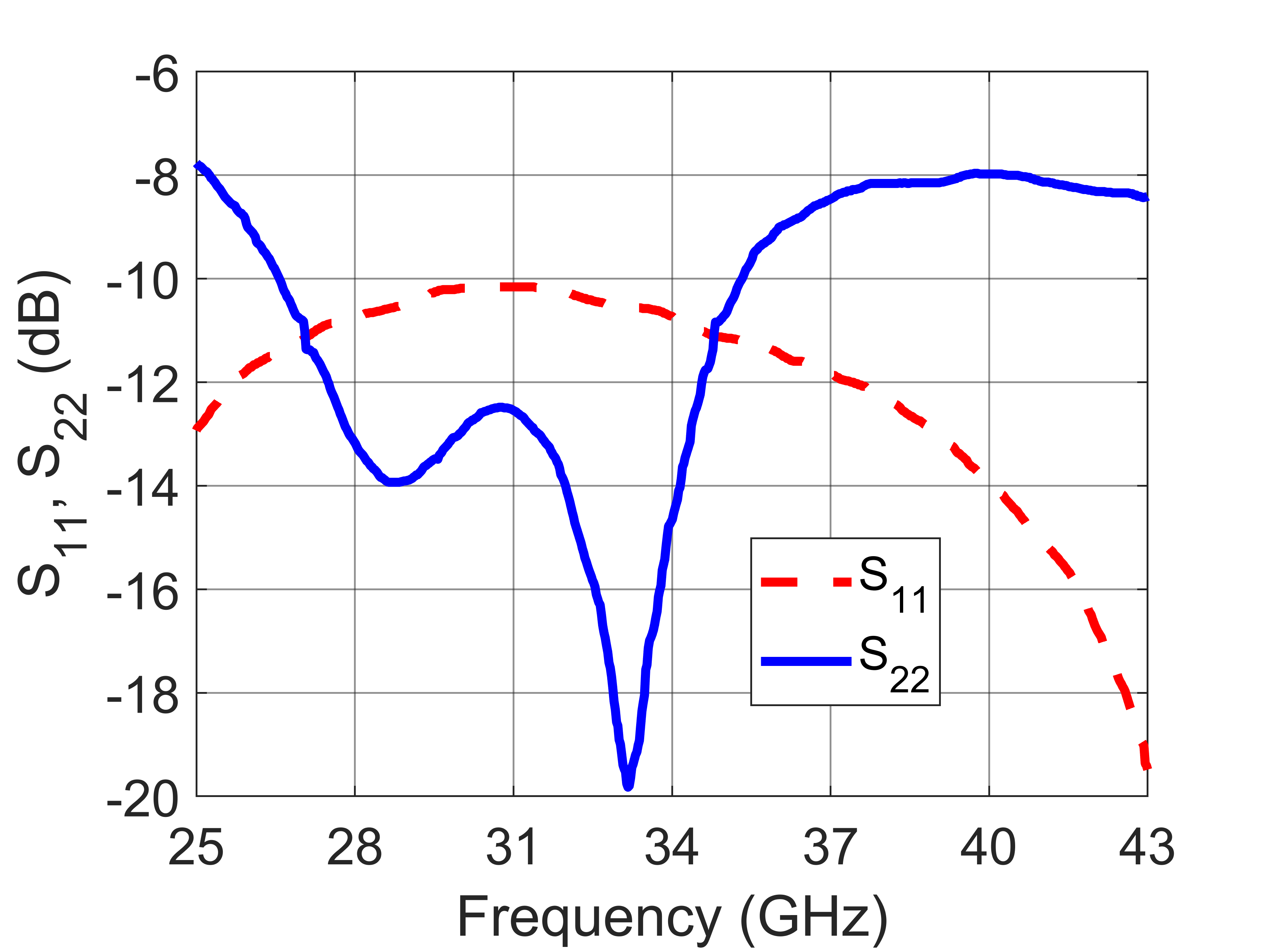}
    \caption{}
  \end{subfigure}
  \caption{Simulated S-parameter for proposed dual-band output matching transformer: (a) $S_{21}$, (b) $S_{11}$ and $S_{22}$.}
  \label{S_Param}
\end{figure}

\begin{figure}[!t]
 \begin{center}
 \includegraphics[width=0.9\columnwidth]{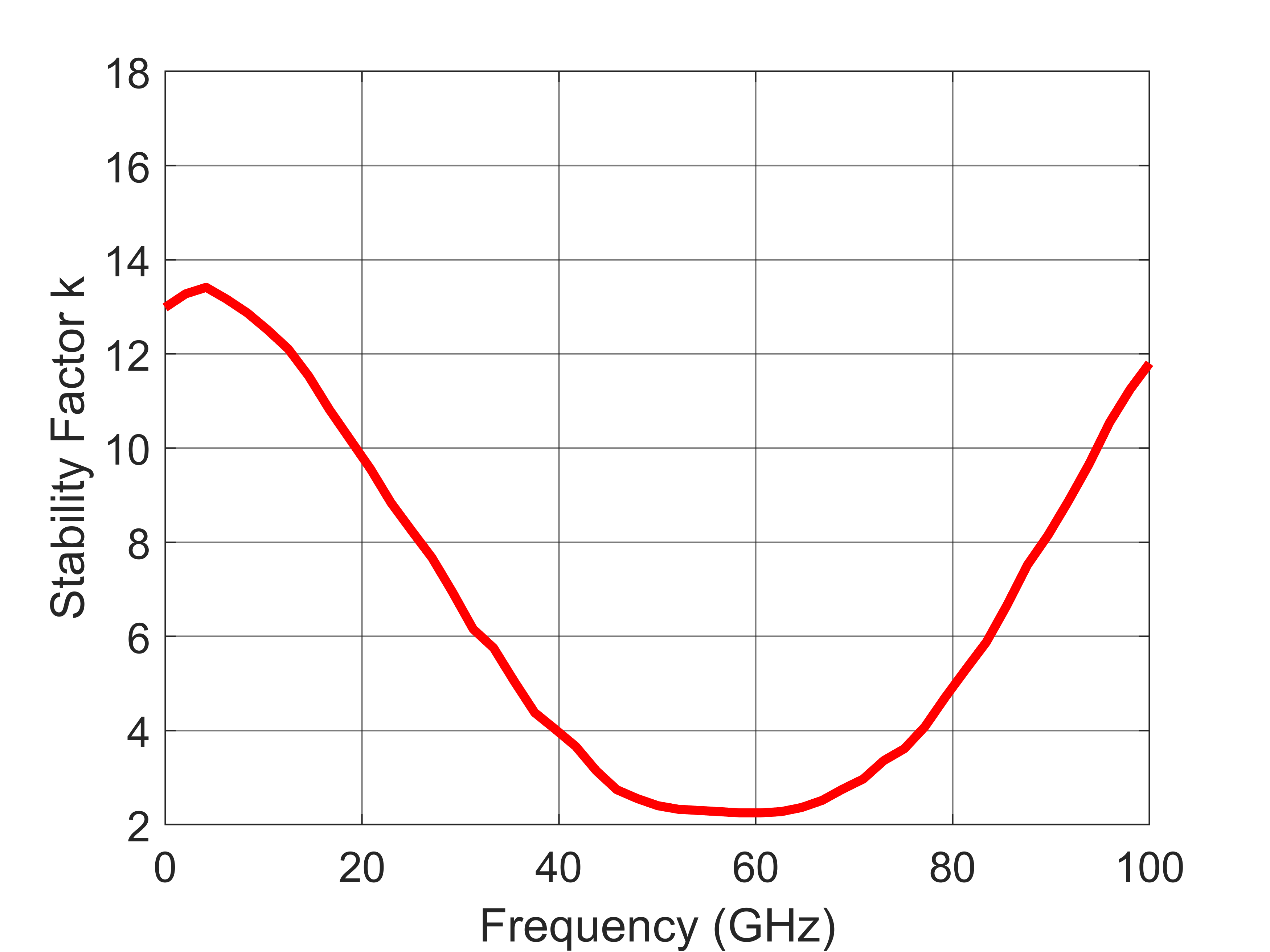}
\caption{Simulated stability K factor of the PA.}\label{K_Factor}
 \end{center}
\end{figure}

Simulated scattering parameters of the PA are shown in Fig. \ref{S_Param}. The small-signal gain is 16.0\,dB at 28\,GHz and 15.5\,dB at 38\,GHz. The inter-band suppression is around 6\,dB which has been limited by low quality factors of the transformers and the resonator. It is expected to achieve higher inter-band suppression using a process with higher passives quality factor. The input return loss is higher than 10\,dB is both bands which indicate a good input impedance matching. The output return loss is 13\,dB at the lower and 8\,dB at the higher band. Furthermore, the stability K factor (Rollet criterion), shown in Fig. \ref{K_Factor}, indicates that the PA is unconditionally stable. 


\subsection{Large-Signal Simulations}

\begin{figure}[t!]
  \centering
  \begin{subfigure}[b]{0.9\linewidth}
    \includegraphics[width=\linewidth]{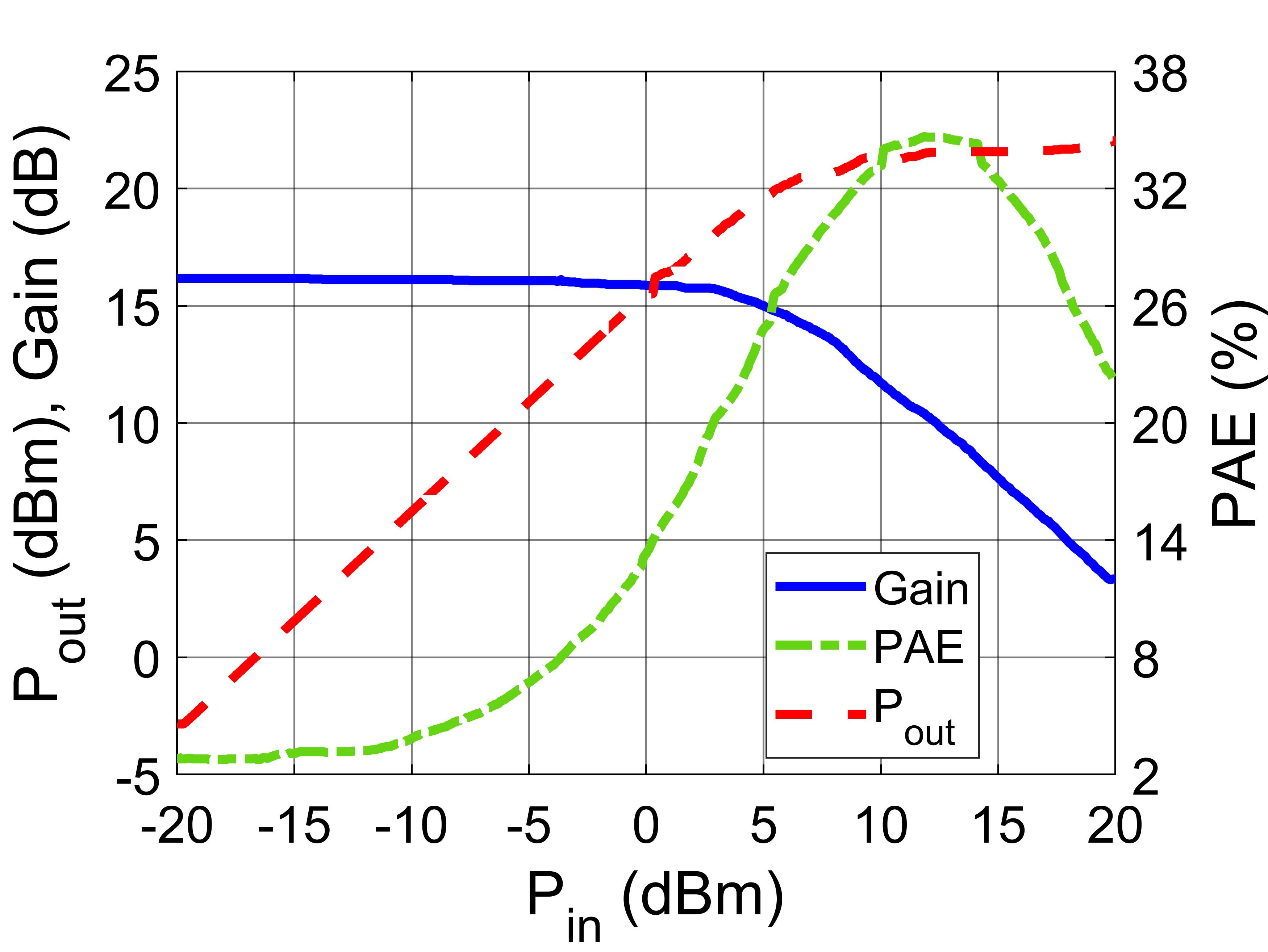}
    \caption{}
  \end{subfigure}
  \begin{subfigure}[b]{0.9\linewidth}
    \includegraphics[width=\linewidth]{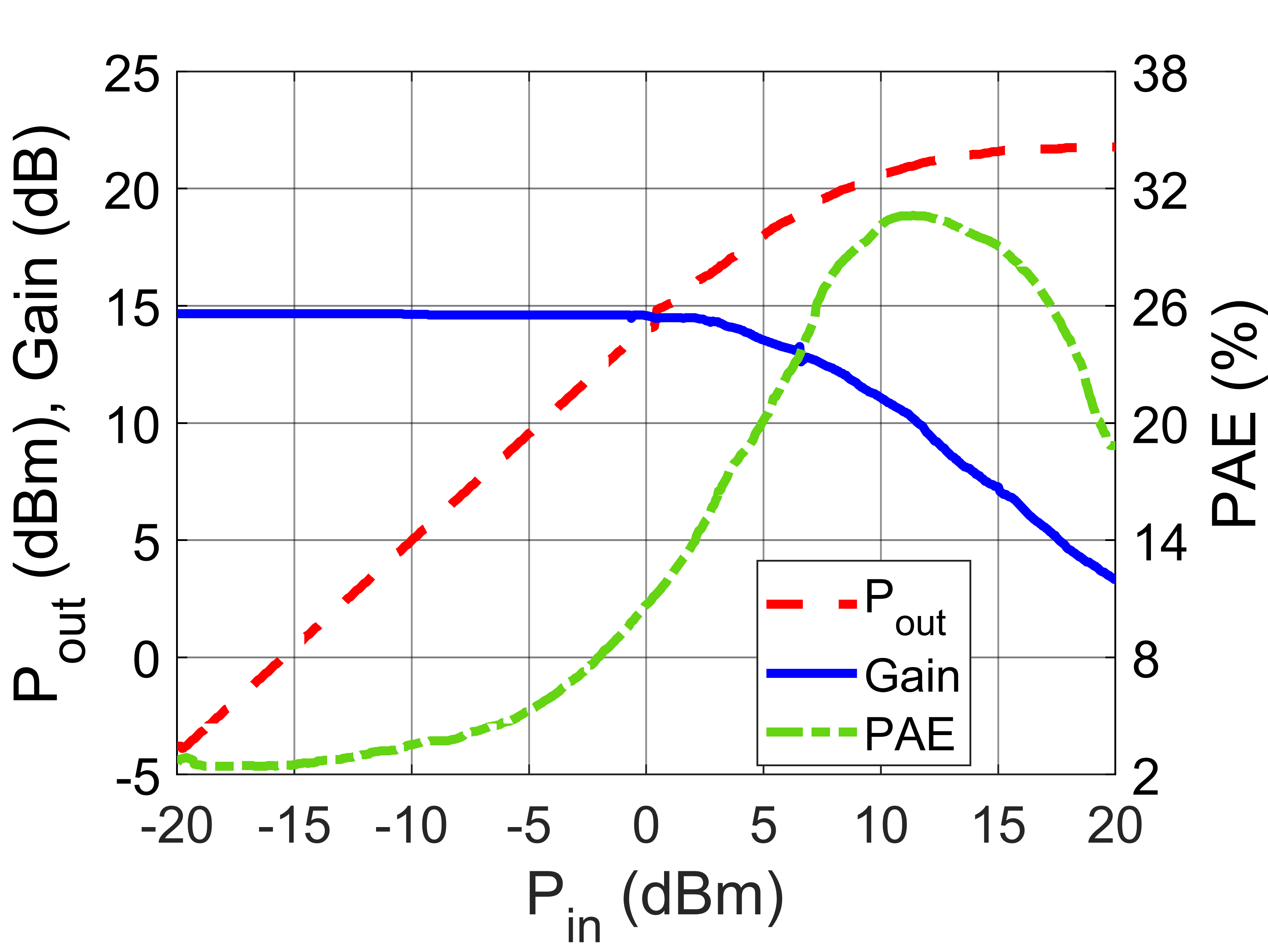}
    \caption{}
  \end{subfigure}
  \caption{Simulated saturated output power, PAE, and gain versus input power at: (a) 28\,GHz (b) 38\,GHz.}
  \label{Large_Signal_28_38}
\end{figure}

Simulations results for the large-signal performance of the PA are shown in Fig. \ref{Large_Signal_28_38}. The saturated output power $\rm P_{sat}$ is 22.6\,dBm at 28\,GHz and 22.0\,dBm at 38\,GHz. The output-input characteristics indicate that the output power is saturated at around 4-dB gain compression. The output power at 1-dB gain compression $\rm P_{1dB}$ is 19.8/20.0\,dBm at 28/38\,GHz, about 2--3\,dB lower than the saturated power. Furthermore, the maximum PAE reads 33/32\% at 28/38\,GHz.


\subsection{Performance Comparison}
A performance comparison of the designed dual-band PA with state-of-the-art mm-wave CMOS PAs operating in similar frequency bands is presented in Table \ref{comparison}. We have indicated the references which are based on simulation results \cite{mayeda2020highly, mayeda2019effective, moret201728ghz} with an asterisk symbol to make a fair comparison. The saturated output power $\rm P_{sat}$ of the PA in this work is over 3\,dB higher than the PAs based on simulations results, while $\rm PAE_{max}$ is also higher than these works. Furthermore, $\rm P_{sat}$ is higher than that of the fabricated PAs except for \cite{daneshgar2020high} with single-band operation, 37\% higher supply voltage, and 6$\times$ chip area. $\rm PAE_{max}$ is competitive with state-of-the-art even though it is challenging to achieve high efficiency for such a high power level. Furthermore, the proposed inter-band suppression technique for the dual-band PA has not been considered in the literature.


\begin{table*}[!t]\footnotesize
	\renewcommand{\arraystretch}{1.5}
	\setlength{\tabcolsep}{5pt}
	\caption{Performance comparison with mm-wave CMOS PAs.}
	\label{comparison}
	\begin{minipage}{7in}
		\centering
		\begin{tabular}{c|c|c|c|c|c|c|c|c}
			\hline
			&f (GHz) &$\mathrm{P_{1dB}}$ (dBm)& $\mathrm{P_{sat}}$ (dBm)& $\mathrm{PAE_{max}}$ (\%) & Gain (dB)& $\mathrm{V_{DD}}$ (V) & Area (mm$\rm ^2$)& Process \\
			\hline\hline
    	\cite{ding202228} & 28/38  & 17.8/15.8 &  18.8/17.0 & 25.0/17.5 & 25.7/21.4 & 3.3 & 0.94 & 250-nm BiCMOS \\
			
         \cite{lee2021millimeter} & 26.5/37 & 18.7/18.6 & 20.2/19.1 & 33.6/32.0 & 13.0/10.7 & NA & 0.11 &28-nm CMOS \\

            \cite{xu202128} & 28/38 & 15.7/15.2 & 19.4/19.2& 26.4/25.1& 31.7/29.2 & 2.7 & 0.26 &22-nm FD-SOI CMOS \\

            \cite{mayeda2020highly}$^{\mathrm{*}}$ & 30/40 & 13.7/13.6  &  17.3/16.4 & 29.2/22.7 & 14.9/13.3 & NA &  NA & 22-nm FD-SOI CMOS \\
			
            \cite{mayeda2019effective}$^{\mathrm{*}}$ & 24/28 & 14.0/15.4  &  17.8/17.3 & 32.0/29.2 & 26.5/22.1 & NA & NA & 22-nm FD-SOI CMOS \\

            \cite{moret201728ghz}$^{\mathrm{*}}$ & 28 & 15.4   & 18.7 & 12.4 & 17.5 & 2.0 &   0.66 & 28-nm CMOS \\

             \cite{shakib2016highly} & 30 & 13.2 &  14.0 & 35.5 & 15.7 & 1.0 & 0.16$^{\mathrm{**}}$& 28-nm CMOS\\

             \cite{dasgupta201926} & 39 & 21.5 &  26.0 & 27.7 & 38 & 2.2 & 2.96 & 28-nm CMOS\\

             \cite{daneshgar2020high} & 39 & 16.6 &  18.3 & 29.1 & 12.5 & 1.8 & 0.48 & 16-nm FinFET CMOS\\

             \cite{tang2021design} & 28 & 18.2 &  19.4 &34.3 & 32 & 2.4 & 0.10 & 22-nm FD-SOI CMOS \\

            \cite{manente202028} & 28 & 20.7 &  21.5 & 26.0 & 20.4 & 1.8 & 0.5& 28-nm CMOS\\	
						
			 \cite{yu201828} & 28 &  17.4  &  18.0 & 26.7 & 19.0 & 1.0 & 0.16 & 65-nm CMOS \\

            This Work$^{\mathrm{*}}$ & 28/38 & 19.8/20.0 & 22.6/22.0 & 33.0/32.0 & 16.0/15.5 &  1.6 &0.45 & 22-nm FD-SOI CMOS \\
			\hline
			\multicolumn{3}{l}{$^{\mathrm{*}}$Simulation results}\,{$^{\mathrm{**}}$Active area}
		\end{tabular}
	\end{minipage}
\end{table*}

\section{Conclusion}
We presented a dual-band transformer network with inter-band suppression which is enabled by a resonator in the center-tap of the transformer. A theory was developed to provide insights into the circuit operation, design guidelines, and impacts of imperfect on-chip circuit elements. A proof-of-concept 28/38-GHz power amplifier (PA) was designed using the proposed network in 22-nm Fully-Depleted Silicon-on-Insulator (FD-SOI) CMOS process. The PA achieved state-of-the-art output power through multiple techniques including the transistor stacking, sizing the power cells for 50-$\Omega$ optimum resistance, and four-path parallel-series transformer combining. The inter-band suppression is 6\,dB at 33\,GHz which is limited by quality factor of inductors and transformers. It can be improved, in a future research, through applying the inter-band suppression in the input transformers, using a modified resonator circuit, and using a process with higher quality factor of inductors.

\section*{Acknowledgment}

The authors would like to thank Global Foundries for the PDK support.

\end{document}